\documentclass{aa}

\usepackage{soul}
\usepackage[T1]{fontenc}
\usepackage{txfonts}
\usepackage{graphicx}	
\usepackage{amsmath}	
\usepackage{amssymb}	
\usepackage{bm}
\usepackage[hidelinks,colorlinks=true,linkcolor=blue,citecolor=blue]{hyperref}
\usepackage[normalem]{ulem}
\usepackage{aas_macros} 

\usepackage[abs]{overpic}
\usepackage{units}

\usepackage[dvipsnames]{xcolor}

\newcommand{\stellarcollapse}{\footnote{\url{https://stellarcollapse.org}}}
\newcommand{\compose}{\footnote{\url{https://compose.obspm.fr}}}

\begin{document} 

\title{The impact of nuclear equations of state on the dynamics and multi-messenger emission of magnetorotational stellar explosions}

\authorrunning{A. Celati et al.}
\titlerunning{The impact of EoS on the dynamics and multi-messenger emission of magnetorotational stellar explosions}

\author{
A. Celati\inst{1,2,3}
\fnmsep\thanks{andrea.celati@unifi.it}
\and 
M. Bugli\inst{4,5,2}
\and 
L. Del Zanna\inst{1,2,3}
 \and
 M. Cusinato\inst{6,4}
\and
 M. Obergaulinger\inst{6,7}
}

\institute{
Dipartimento di Fisica e Astronomia, Universit\`a di Firenze, Largo E. Fermi 2, I-50125 Firenze, Italy
\and
INAF, Osservatorio Astrofisico di Arcetri, Largo E. Fermi 5, I-50125 Firenze, Italy
\and
INFN, Sezione di Firenze, Via G. Sansone 1, I-50019 Sesto Fiorentino (FI), Italy
\and
Université Paris-Saclay, Université Paris Cité, CEA, CNRS, AIM, F-91191 Gif-sur-Yvette, France
\and
Institut d’Astrophysique de Paris, UMR 7095, CNRS \& Sorbonne Universit\'e, F-75014 Paris, France
\and
Departament d'Astronomia i Astrofisica, Universitat de València,  Av.~Vicent Andrés Estellés 19, 46100, Burjassot (Val\`encia), Spain
\and
Observatori Astronòmic, Universitat de València, 46980 Paterna (València), Spain
}

\date{Received XXX; accepted YYY}

  \abstract
   {The gravitational collapse of massive stars at the end of their life leads to powerful supernova explosions that produce new stellar-sized compact objects, regulate the dynamics of their host galaxies, and produce new heavy elements that contribute to the cosmic chemical evolution. In presence of fast rotation and strong magnetic fields, such explosions reach extremely high energies that can explain sources such as hypernovae and long gamma-ray bursts, which are the most violent transients observable in the Universe.}
   {We test the impact of variations in the uncertain nuclear equation of state (EoS) on the dynamics of magnetorotational explosions and the resulting multi-messenger emission, including neutrinos and gravitational waves. In particular, differences in the stiffness, composition, and finite-temperature behavior of the EoS can significantly affect the collapse, bounce, and jet-launching phases. }
   {We use the \texttt{Aenus-Alcar} code, employing relativistic magnetohydrodynamics (RMHD), two-moment neutrino transport, neutrino-matter interactions, and general-relativistic corrections, to perform axisymmetric models using different EoSs. All simulations start from the same initial condition, using a standard pre-supernova model with solar metallicity and a zero-age main sequence mass of $\unit[20]{M_\odot}$, endowed with a dipolar magnetic field configuration and a shellular rotation profile.}
   {Variations in the stiffness, nuclear interactions, and treatment of nuclei among the different EoSs lead to significant differences in the explosion dynamics, especially in the bounce time, the proto-neutron star properties, the mass of the ejected material, and the associated multi-messenger signals.
  }
   {Our results demonstrate that the multimessenger signatures of magnetorotational core-collapse supernovae (CCSNe) are sensitive not only to the cold stiffness of the EoS, but also to its thermal and compositional properties. These findings highlight the importance of combining gravitational-wave and neutrino observations to constrain the microphysics of dense matter and the explosion mechanism of rapidly rotating CCSNe.}


   \maketitle

\section{Introduction}

One of the most energetic phenomena in the Universe is the gravitational collapse of a massive star, which releases a binding energy on the order of $10^{53}\,\mathrm{erg}$. While about 99\% of this energy is carried away by neutrinos emitted during the collapse and the subsequent cooling of the proto-neutron star (PNS)—the compact stellar remnant formed in the innermost region of the collapsing core—the remaining fraction is sufficient to power the core-bounce shock wave and drive the core-collapse supernova (CCSN) explosion \citep{Janka2017}.

The neutrino-heating mechanism is widely considered the dominant explosion mechanism for CCSNe, in which neutrino interactions with neutrons and protons in the gain region deposit enough energy to revive the stalled shock and overcome the ram pressure of the infalling material (see \citealt{Janka2012}). This mechanism provides a framework capable of reproducing many of the observed light-curve properties of CCSNe \citep[e.g.,][]{Curtis2021}, but it cannot explain the exceptionally high luminosities of superluminous supernovae \citep{nichollSlowlyFadingSuperluminous2013, greinerVeryLuminousMagnetarpowered2015}, unless the strong shock interacts with a dense circumstellar medium \citep{smith2014, Inserra2017}.
Moreover, the neutrino-driven mechanism  produces ejecta with kinetic energies an order of magnitude lower than those inferred for hypernovae \citep{iwamotoHypernovaModelSupernova1998} and long gamma-ray bursts \citep[LGRBs,][]{ soderbergRelativisticEjectaXray2006, droutFIRSTSYSTEMATICSTUDY2011}.

A very promising candidate to explain the most energetic transients is the magnetorotational mechanism, in which an intense magnetic field efficiently extracts rotational energy from the PNS through magnetic braking, thereby powering extreme explosions \citep{BisnovatyiKogan1970,BisnovatyiKogan1980MagnetorotationalModel,LeBlancWilson1970}. 
Modern multidimensional simulations have further explored this scenario and confirmed its viability under favorable conditions 
\citep{wintelerMAGNETOROTATIONALLYDRIVENSUPERNOVAE2012,mostaMAGNETOROTATIONALCORECOLLAPSESUPERNOVAE2014,Kuroda2020,bugliThreedimensionalCorecollapseSupernovae2021,Shibagaki2024}. It is evident that both rapid rotation of the PNS and a strong magnetic field are required to trigger such explosions, which explains their rarity. However, it remains debated how such a combination can be produced during the collapse of a massive star.

Uncertainties in progenitor magnetic fields \citep{WoosleyHeger2006,aguilera-denaRelatedProgenitorModels2018} and simplified dynamo prescriptions \citep{spruitDynamoActionDifferential2002,fullerSlowingSpinsStellar2019} still limit CCSN models.
During the progenitor's evolution, various mechanisms may amplify magnetic fields, but it remains unclear whether they can produce magnetar-level strengths ($\sim10^{15}\,\mathrm{G}$) after collapse while preserving much of the angular momentum in the core. 
The Tayler-Spruit dynamo \citep{spruitDynamoActionDifferential2002} and fossil-field scenarios \citep{Shultz2018} both tend to slow down stellar rotation through angular-momentum transport \citep{AngularMomentumTransport} or magnetic braking, and stellar mergers also yield slowly rotating remnants \citep{schneiderStellarMergersOrigin2019}. Amplification within the PNS therefore provides a promising alternative, with convection \citep{raynaudMagnetarFormationConvective2020} and magnetorotational instability \citep{BalbusHawley1998,Akiyama2003,SemiglobalSimulationsMagnetorotational,GuiletMuellerJanka2015,reboul-salzeGlobalModelMagnetorotational2021} capable of generating strong large-scale fields largely independent of the initial configuration \citep{raynaudMagnetarFormationConvective2020,reboul-salzeGlobalModelMagnetorotational2021}.
The saturated magnetic field within the PNS typically exhibits a complex topology, with only 2-3\% of the magnetic energy in the dipole component and an axis nearly orthogonal to rotation, motivating simulations exploring multipolar and misaligned configurations \citep{Bugli2020,bugliThreedimensionalCorecollapseSupernovae2021,bugliThreedimensionalCorecollapseSupernovae2023,reichert2024}.

An interesting aspect of CCSNe is that they are multi-messenger sources. During gravitational collapse, the neutronization process converts electrons and protons into neutrons and electron-type neutrinos, which escape the core and lead to a sharp decrease of the electron lepton number. Owing to the highly degenerate conditions in the collapsing core, positron production is strongly suppressed at this stage. Positrons can instead be produced later on, in the hotter and less degenerate outer layers of the newly formed PNS, through other processes. They then interact with neutrons and with electrons producing electron antineutrinos and heavy-lepton neutrinos, respectively \citep{Janka2012}. Typically, muon and tau neutrinos and antineutrinos are treated as a single heavy-lepton species because their interaction rates are very similar under the conditions relevant for CCSN simulations \citep{Janka2017}.
25 electron antineutrinos
have already been observed: namely during Supernova SN~1987A by the Super-Kamiokande neutrino observatory \citep{Hirata1987SN1987A}. Current detectors for electron antineutrinos include ORCA in the KM3NeT Collaboration \citep{AdrianMartinez2016KM3NeTORCA}, IceCube \citep{Aartsen2017IceCube} and Super-Kamiokande \citep{Abe2022SuperKamiokandeDetector}, while future observatories such as DUNE \citep{Abi2020DUNE} and DarkSide \citep{Aalseth2018DarkSide20k} will be sensitive to mostly electron neutrinos and all neutrino species, respectively.

In addition to being intense sources of neutrinos, CCSNe are also expected to emit gravitational waves \citep[GWs,][]{Ott2009,Murphy2009,Yakunin2010,powellGravitationalWavesCorecollapse2025}. While neutrinos carry direct information about the thermodynamic and weak-interaction processes occurring deep inside the PNS, GWs provide a complementary probe of its hydrodynamic and rotational behavior. 
Moreover, CCSNe are multimessenger sources both in fully three-dimensional and in axisymmetric simulations. In axisymmetry, the GW signal contains only the plus polarization mode and vanishes along the polar direction, while emission at intermediate angles is still present. In contrast, full 3D models produce both polarization modes and yield a non-vanishing signal for all observer directions. This highlights an important difference between GW and neutrino emission: while neutrinos are emitted in all directions in both 2D and 3D, GW observables are strongly affected by the dimensionality of the model.
The GW signal originates from several physical processes: the core bounce in cases where the PNS is significantly deformed by rotation, the subsequent oscillations of the PNS, and convective motions occurring both within the PNS and in the region below the shock wave \citep{Mezzacappa2024}. A long-standing problem concerns identifying which specific regions of the PNS and physical processes contribute most to the GW emission. An example of a study addressing this issue is presented by \cite{cusinatoConvectionSignaturesEarlytime2025}, where the authors use time–space maps to locate the origin of GW emission arising from convection during the early post-bounce phase. 
So far, no GW signal from a CCSN has been directly observed \citep{abacSearchGravitationalWaves2025}, mainly because current detectors are sensitive to such events only within distances up to approximately $100\,\mathrm{kpc}$, even in cases of rapid rotation. GW observations are currently performed by the Advanced LIGO \citep{aasiCharacterizationLIGODetectors2015}, Advanced Virgo \citep{acerneseAdvancedVirgoDetector2015}, and KAGRA \citep{akutsuOverviewKAGRA25Generation2021} interferometers, with future prospects for the Einstein Telescope and Cosmic Explorer to greatly improve detection capabilities.

One of the crucial ingredients in CCSN modeling is the choice of the nuclear equation of state (EoS). The transition from inhomogeneous to homogeneous nuclear matter determines the sudden halt of the inner core’s gravitational collapse through the nuclear interactions described by the EoS. The nuclear force underlying the EoS represents an effective quantum many-body interaction and remains one of the least constrained aspects of fundamental physics, introducing significant uncertainties in astrophysical scenarios involving compact stellar objects. 
The nuclear EoS can be constrained by laboratory experiments \citep[e.g.][and references therein]{lattimerNuclearEquationState2012,oertelEquationsStateSupernovae2017}, by theoretical nuclear physics calculations \citep{hebelerConstraintsNeutronStar2010,hebelerEQUATIONSTATENEUTRON2013,tewsSymmetryParameterConstraints2017}, by astronomical observations of neutron star (NS) masses and radii \citep{lattimerNuclearEquationState2012,nattilaEquationStateConstraints2016,ozelMassesRadiiEquation2016}, and by GW detections \citep{aasiCharacterizationLIGODetectors2015,acerneseAdvancedVirgoDetector2015,akutsuOverviewKAGRA25Generation2021}.
Besides the structure and stability of NSs, the main astrophysical contexts where the nuclear EoS plays a key role are CCSNe and NS mergers. In the latter case, the tidal deformability of NSs depends sensitively on the EoS, affecting the late-inspiral GW signal in a measurable way \citep{bernuzziTidalEffectsBinary2012,bernuzziModelingCompleteGravitational2015,flanaganConstrainingNeutronstarTidal2008,readMeasuringNeutronStar2009}. During the merger, the tidal disruption of a NS in a NS–black hole system produces an abrupt cutoff in the GW signal, which can also be used to constrain EoS properties \citep{readMeasuringNeutronStar2009,vallisneriProspectsGravitationalWaveObservations2000,shibataMergerBlackHole2008}. Moreover, the post-merger remnant of a binary NS collision emits GWs efficiently, with spectral features that can be directly linked to the underlying nuclear EoS \citep{radiceProbingExtremeDensityMatter2017,bausweinMeasuringNeutronStarProperties2012,bausweinRevealingHighdensityEquation2014,bernuzziHowLoudAre2016}. 
While NS mergers have been extensively studied in this context, the impact of the nuclear EoS on CCSNe has been investigated primarily in the framework of neutrino-driven explosions, both in terms of dynamics \citep{Janka2012,suwaIMPORTANCEEQUATIONSTATE2013,Yasin2018EoS,powell2025} and multimessenger emission \citep{marekEquationofstateDependentFeatures2009,Richers2017,eggenbergerandersenEquationofstateDependenceGravitational2021,jakobusGravitationalWavesCore2023,murphyDependenceReconstructedCorecollapse2024}. 
In contrast, the role of the EoS in magnetorotational explosions has received comparatively less attention.

In this work, we present the first study investigating the impact of microphysics on numerical models of magnetorotational CCSNe, by performing a series of axisymmetric simulations that differ only in the choice of the nuclear EoS. We focus on its effects on the explosion dynamics and on the resulting multi-messenger signals. Both the explosion dynamics and the PNS properties are affected by this choice: differences in the microphysical response of matter lead to variations in the explosion efficiency and timescale, as well as in the rotational properties of the PNS, thereby directly influencing the efficiency of the magnetorotational mechanism itself.

The structure of the paper is as follows. Sect.~\ref{sect:2} describes the numerical setup, initial conditions, and the EoSs employed, Sect.~\ref{sec:results} presents the main results of our simulations, while Sect.~\ref{sect:conclusion} summarizes our conclusions and outlook.

\section{Physical and numerical setup}
 \label{sect:2}
 The simulations described in this work were performed using the relativistic magnetohydrodynamics (RMHD) \texttt{Aenus-Alcar} code \citep{Just2015}, assuming axisymmetry and employing a pseudo-Newtonian gravitational potential which incorporates general-relativistic corrections \citep[Case A of ][]{Marek2006EffectivePotential}. The code solves the \(\nu\)-RMHD set of equations, i.e., the coupled system of RMHD and two-moment (M1) neutrino transport, in which the equations for the neutrino energy and momentum densities \citep{MunierWeaver1986,CernohorskyvanWeert1992,Cardall2013} are closed by a local algebraic pressure tensor, in our case the based on the maximum-entropy Eddington factor \citep{CernohorskyBludman_1994}, as described in \cite{Obergaulinger2020}. For further details see Appendix~\ref{App:eqs}. 

Transport equations are discretized using a finite-volume scheme. To ensure the solenoidal condition of the magnetic field ($\nabla \cdot \vec{B} = 0$), the upwind constrained transport (UCT) method is employed \citep{LondrilloDelZanna2004,MignoneDelZanna2021}.  
High-resolution shock-capturing properties are achieved by combining the monotonicity-preserving MP5 reconstruction scheme \citep{sureshAccurateMonotonicityPreservingSchemes1997} with the two-wave HLL Riemann solver \citep{HartenLaxVanLeer1983}, as in \cite{DelZanna2007}.

The simulations are carried out on a two-dimensional spherical grid in the $(r,\theta)$ plane. The angular domain spans the full polar range, $\theta \in [0,\pi]$, and is discretized with 128 uniformly spaced zones. Reflection boundary conditions are imposed on the symmetry axis.
The radial direction is discretized using a uniformly spaced grid with resolution $\Delta r = 5\times10^{4}\,\mathrm{cm}$ up to a transition radius $R_0 = 2 \times10^{6}\,\mathrm{cm}$. Beyond $R_0$, a logarithmically stretched grid is used to extend the computational domain up to an outer radius of $R_{\text{out}} = 1.69 \times 10^9\, \mathrm{cm}$, discretized with 320 radial zones. The choice of the transition radius ensures that the aspect ratio of the grid cells, defined as $\Delta r /(r\Delta\theta)$, remains approximately unity throughout most of the domain, thereby producing quasi-square cells. Near the center, where $\Delta r$ is constant, an angular coarsening scheme is adopted in order to avoid prohibitively small timesteps imposed by the Courant-Friedrichs-Lewy condition while approximately preserving the cell aspect ratio.
The neutrino transport employs a spectral resolution of 12 energy bins, which are logarithmically spaced to better capture the energy dependence of neutrino interactions between $\epsilon_{\mathrm{MIN}}=1\,\mathrm{MeV}$ and $\epsilon_{\mathrm{MAX}}=300\,\mathrm{MeV}$.

\subsection{EoS properties}
\label{sub:EoS detail}

To describe matter at densities above the threshold of $\rho = 10^8\,\mathrm{g/cm^3}$, we adopt for each run one of four selected nuclear EoSs, namely SFHo \citep{Steiner2013}, LS220 \citep{Lattimer1991}, DD2 \citep{hempelStatisticalModelComplete2010}, and SLy4 \citep{schneiderOpensourceNuclearEquation2017}. These EoSs are described in detail in this section, while a summary of the main parameters are reported in Table~\ref{tab:EoS}. The corresponding mass–radius relations for cold, beta-equilibrated, and static NS configurations are reported in Appendix~\ref{App:m-r}. Densities below the threshold are always treated with a low-density EoS that includes contributions from leptons, photons, and baryons \citep{RamppJanka2002}. The latter include free neutrons, protons and heavy nuclei. In the flashing scheme, the latter are approximated by a composition of pure \textsuperscript{28}Si for temperatures below 0.44~MeV and pure \textsuperscript{56}Ni for higher temperatures.

All EoSs are provided in tabulated form as functions of density, temperature, and electron fraction. SFHo, LS220, and DD2 tables are taken from the \texttt{stellarcollapse.org} repository\stellarcollapse,
while SLy4 is obtained from \texttt{CompOSE}\compose.
The choice of nuclear EoS directly impacts macroscopic and observable properties of NS and PNS, including their mass–radius relation, thermal structure, and neutrino emission, which will be discussed in detail in Section~\ref{sec:results}.

\begin{table}
\renewcommand{\arraystretch}{1.2}
\begin{center}
\caption{\small Nuclear matter and NS properties of the adopted EoSs.}

\label{tab:EoS}
 \begin{tabular}{lcccc}
\textbf{Property} & \textbf{SLy4} & \textbf{SFHo} & \textbf{LS220} & \textbf{DD2} \\
\hline
Model type & Skyrme & RMF & Skyrme & RMF \\
Composition & SNA & NSE & SNA & NSE \\
\( K \) [MeV]  & 230 & 245 & 220 & 242 \\
\( S \) [MeV]  & 32.04 & 31.57 & 28.61 & 31.67 \\
\( L \) [MeV]  & 46.0 & 47.10  & 73.80  & 55.03 \\
$\mathrm{R}_{1.4}$  [km] & 11.68 & 11.92 & 12.66 & 13.26 \\
$\mathrm{M}_{max}$ [$M_{\odot}$] & 2.05 & 2.06& 2.06& 2.42\\
\hline
\end{tabular}
\tablefoot{
The nuclear incompressibility ($K$) is the coefficient of the second-order term in the power-series expansion of the binding energy per baryon around nuclear saturation density. $S$ is the symmetry energy of the Bethe-Weizsäcker mass formula evaluated at saturation density, and $L$ is its density slope. These three parameters characterize the stiffness of the EoSs around saturation density. $R_{1.4}$ denotes the radius of a cold, beta-equilibrated neutron star with a gravitational mass of $1.4\,M_\odot$, while $\mathrm{M}_{\mathrm{max}}$ denotes the maximum gravitational mass supported by a cold, beta-equilibrated neutron star.
}
\end{center}
\end{table}

An important microphysical distinction among the adopted EoSs concerns the treatment of heavy nuclei in the sub-nuclear density regime. SFHo and DD2 are based on a Nuclear Statistical Equilibrium (NSE) description \citep{hempelStatisticalModelComplete2010}, in which a full ensemble of nuclear species is included. In the final tabulated form, the resulting composition is represented in terms of average nuclear properties (e.g. mean mass and charge numbers). This differs from a classical Single Nucleus Approximation \citep[SNA,][]{burrowsAccuracySinglenucleusApproximation1984a}, as used in LS220 and SLy4, where a single representative heavy nucleus is selected based on the liquid-drop model.

In addition to the composition treatment, the underlying nuclear interaction differs between Skyrme-type (LS220 and SLy4) and relativistic mean field (RMF)-type (SFHo and DD2) models. RMF approaches are based on meson-exchange interactions within a covariant framework, whereas Skyrme models employ non-relativistic effective interactions that typically provide a more phenomenological description of nuclear matter. These differences affect the symmetry energy, $S$, its density dependence, and the pressure around saturation density, thereby influencing the stiffness of the EoS and the resulting NS structure.

The transition from inhomogeneous nuclear matter (nuclei embedded in a nucleon gas) to homogeneous nuclear matter (uniform nucleon fluid) also depends on the underlying nuclear interaction and nuclei treatment.
In NSE-based RMF EoSs (SFHo and DD2), nuclei gradually dissolve as the density approaches saturation, leading to a smooth change in composition and thermodynamic quantities such as pressure, chemical potentials, and entropy.
In contrast, in SNA-based Skyrme EoSs (LS220 and SLy4) nuclei dissolve into uniform matter at a defined density threshold based on energetics of the liquid-drop model, producing a sharper transition compared to the smooth RMF treatment.

Throughout this work, we often refer to one nuclear EoS as being ``stiffer'' or ``softer'' than another. Here, this terminology refers to the radius of a cold, non-rotating NS with mass $1.4\,M_\odot$
(see Appendix~\ref{App:m-r}). In the hot, lepton-rich PNS formed after core collapse, matter reaches and exceeds nuclear saturation density, typically taken as $\rho_{\mathrm{sat}} \simeq 2.7 \times 10^{14}\,\mathrm{g,cm^{-3}}$, and central temperatures $T \sim 5$--$30,\mathrm{MeV}$ \citep{Pons1999,Huedepohl2010,Fischer2010}. Under these conditions, the effective stiffness depends on both the zero-temperature nuclear interaction and thermal/compositional effects. For reference, the effective stiffness hierarchy for a cold, spherically symmetric, non-rotating NS is: \texttt{DD2} $>$ \texttt{LS220} $>$ \texttt{SFHo} $>$ \texttt{SLy4}, where ``$>$'' indicates a stiffer EoS. This ranking may vary outside the stated density/temperature ranges, and composition and lepton fraction further influence the PNS behavior. By including EoSs with different thermal behavior, composition, nuclei treatment, and phase transition properties, our work captures the main microphysical factors differentiating the space of possible EoSs in realistic core-collapse simulations.
We stress that the stiffness hierarchy quoted above refers to cold NS configurations and does not uniquely determine the effective behavior of the EoS under hot, lepton-rich PNS conditions.
All the EoSs employed in this study are compatible with the existence of NSs with maximum masses of at least $\sim 2\,M_\odot$, as required by current observational constraints from massive pulsars. However, they span a range of radii and tidal deformabilities, reflecting remaining theoretical uncertainties in the nuclear EoS.
In particular, LS220 is disfavored by nuclear physics constraints on $S$ and its slope, as discussed in \citet{tewsSymmetryParameterConstraints2017}, due to its comparatively low value of $S$. In contrast, the other three EoSs considered here are more consistent with the combined set of current astrophysical and nuclear constraints.

Below we provide a detailed description of the EoSs employed in this work.

SFHo: This EoS is based on a covariant Lagrangian within the Walecka model framework, where nucleons interact via the exchange of mesons $\sigma$, $\rho$, and $\omega$ in the RMF approximation. The non-linear Walecka model exhibits only small variation in the isospin sector, with the $S$ primarily controlled by the coupling between nucleons and the $\rho$ meson. Additional terms such as $\rho^4$ and $\sigma^2 \rho^2$ are included to increase flexibility, and the sound speed is automatically constrained to remain subluminal. Model parameters are fitted to reproduce the observed mass-radius relation \citep{steinerEquationStateObserved2010}, predicting charge radii and binding energies of $^{208}\mathrm{Pb}$ and $^{90}\mathrm{Zr}$ within 2\% of experimental values \citep[][]{Steiner2013,hempelStatisticalModelComplete2010}.

DD2: Like SFHo, DD2 is RMF-based with $\sigma$, $\rho$, and $\omega$ meson interactions, but with a stiffer parametrization, resulting in higher pressures at a given density \citep{typelCompositionThermodynamicsNuclear2010,hempelStatisticalModelComplete2010,hempelNEWEQUATIONSSTATE2012}. Thermal and compositional effects in DD2 are handled similarly to SFHo, with NSE used to model heavy nuclei.

LS220: This EoS is based on a compressible liquid drop model for nuclei, including surface and Coulomb lattice terms. The nucleon-nucleon interaction is described by a Skyrme-type model, represented by a local Hamiltonian depending on proton and neutron densities and their kinetic energies \citep{Lattimer1991}.

SLy4: This EoS is derived from a Skyrme-type effective nucleon-nucleon interaction, calibrated to reproduce both finite nuclei and infinite nuclear matter properties \citep{schneiderOpensourceNuclearEquation2017}. It includes density-dependent terms in $S$, which determine its behavior in neutron-rich matter and influence the pressure around and above nuclear saturation density. Like LS220, nuclei are modeled within the SNA, capturing a single representative nucleus.

\subsection{Neutrino transport and neutrino-matter interactions}
Neutrino transport is treated within a two-moment framework, in which the evolution equations for the neutrino energy and momentum densities are solved and closed by means of an algebraic maximum-entropy closure for the Eddington factor \citep{CernohorskyBludman_1994}.
The closure relation provides the second angular moment of the neutrino distribution as a function of the lower-order moments. In the chosen comoving-frame formulation, the neutrino transport equations are solved retaining terms up to $O(v/c)$ \citep{Cardall2013}. The momentum equation involves the third angular moment, which is reconstructed using the closure prescription described in \cite{Just2015,Vaytet2011Multigroup}.

General-relativistic corrections are included within the pseudo-Newtonian framework through additional terms involving the lapse function $\alpha$. These terms account for gravitational redshift and time-dilation effects in the neutrino transport equations. Their numerical treatment follows an approach formally analogous to that adopted for the velocity-dependent terms in the two-moment scheme, as discussed in \cite{Just2015} and \cite{Obergaulinger2020}. Further details on the neutrino transport implementation are provided in Appendix~\ref{sub:neu_trans}.

Neutrino-matter interactions in our simulations build upon the basic set of reactions included in \cite{Obergaulinger2014}, which accounts for nucleonic absorption, emission, and scattering, nuclear absorption, emission, and scattering, and inelastic scattering off electrons. The set of neutrino-matter interactions used in the present simulations includes pair processes, specifically electron–positron annihilation and nucleonic bremsstrahlung, following the prescriptions of \cite{Pons1998} and \cite{Hannestad1998}, respectively.

\subsection{Progenitor model}

Our simulations follow the evolution of the stellar model $\texttt{s20}$ described in \cite{WoosleyHeger2007}. This pre-supernova model has solar metallicity and a zero-age main sequence (ZAMS) mass of $\unit[20]{M_\odot}$. The progenitor features a non-convective iron core with mass $M_{\mathrm{Fe}} \simeq 1.54\, M_\odot$ and radius $R_{\mathrm{Fe}} \simeq 1.65 \times 10^8\, \mathrm{cm}$, while the oxygen shell has a mass $M_{\mathrm{O}} \simeq 1.82\, M_\odot$ and extends to a radius $R_{\mathrm{O}} \simeq 2.55 \times 10^8\, \mathrm{cm}$. A drop in the density profile is observed at the location of the oxygen shell. This indicates a transition in the progenitor structure, which in many CCSN models may approximately correspond to the location of the mass cut, i.e., the boundary between the material that eventually becomes part of the compact remnant and the material that may be ejected.

The pre-supernova model \texttt{s20} is provided as a series of radial stellar profiles; density, temperature, and electron fraction are interpolated onto our computational mesh, preserving spherical symmetry. The nuclear composition of the progenitor is not retained explicitly; instead, it is replaced by the simplified Si/Ni mixture prescribed by the low-density EoS treatment, according to the local thermodynamic conditions. The EoS adopted in the progenitor evolution differs from the nuclear EoSs employed in the CCSNe simulations. After mapping, all other thermodynamic quantities are recomputed consistently using the EoS employed in the simulations.

The stellar model $\mathrm{s20}$ does not include rotation nor magnetic fields; therefore, they both must be superimposed on the other initial conditions. We begin by discussing the magnetic field configuration. A number of numerical studies adopt the following prescription for the azimuthal component of the vector potential to initialize a modified dipolar magnetic field configuration \citep{Suwa2007}
\begin{equation}
    A_\phi^{\text{dip}} = \frac{B_0}{2} \frac{r_0^3}{r^3 + r_0^3} r \sin\theta,
\end{equation}
where $B_0 = 3.14 \times 10^{12}\,\mathrm{G}$ is the characteristic magnetic field strength at the stellar center, and 
$r_0 = 10^3\,\mathrm{km} = 10^8\,\mathrm{cm}$, sets the characteristic radial scale of the dipole field.
This setup produces a purely poloidal magnetic field which is approximately uniform and aligned with the symmetry axis for $r \lesssim r_0$. For $r \gtrsim r_0$, the field transitions smoothly to a dipolar configuration, decaying as $\sim r^{-3}$.
Additionally, we impose the following shellular profile of the angular velocity
\begin{equation}
    \Omega = \Omega_0\frac{r_0^2}{r^2+r_0^2},
\end{equation}
where we set $\Omega_0=1\,\mathrm{rad}/\mathrm{s}$.
With this profile, there is rigid rotation up to \( r_0 \).  Beyond this region, the angular velocity decreases as $\Omega \propto r^{-2}$, corresponding to a radially constant specific angular momentum. The choice of the same characteristic radius $r_0$ as in the previous equation for the magnetic field places the transition between flat and $r$-dependent profiles in the same region for both quantities.
These values are intended to represent rapidly rotating and strongly magnetized progenitors typically adopted in magnetorotational CCSN studies \citep{Bugli2020, obergaulingerCoreCollapseMagnetic2018}, and correspond to the upper end of (or exceed) the ranges predicted by current stellar evolution models including angular momentum transport and magnetic braking.

\begin{table*}[t]
\centering
\caption{\small Summary of PNS, ejecta, neutrino, and GW properties for the different CCSN models  at $t=200$~ms and $t=400$~ms post-bounce.}
\label{tab:summary_ccsn}

\begin{tabular}{lcccc}
\hline
\textbf{Quantity} & \textbf{SFHo} & \textbf{LS220} & \textbf{DD2} & \textbf{SLy4} \\
\hline
$t~[\mathrm{ms}]$ & $200$ / $400$ & $200$ / $400$ & $200$ / $400$ & $200$ / $400$ \\
$t_{\mathrm{exp}}~[\mathrm{ms}]$ & $58.00$ & $80.00$ & $63.00$ & $53.00$\\
$t_{\mathrm{bounce}}~[\mathrm{ms}]$ & $270.00$ & $338.00$ & $270.00$ & $249.00$\\
$R_\mathrm{pns,north}~[\mathrm{km}]$ & $48.81$ / $37.58$ & $53.68$ / $42.33$ & $51.19$ / $40.36$ & $52.42$ / $41.33$ \\
$R_\mathrm{pns,eq}~[\mathrm{km}]$ & $57.65$ / $48.81$ & $66.49$ / $47.67$ & $60.46$ / $52.42$ & $61.91$ / $53.68$ \\
$M_\mathrm{pns}~[\mathrm{M_\odot}]$ & $1.73$ / $1.80$ & $1.84$ / $1.96$ & $1.73$ / $1.81$ & $1.71$ / $1.79$ \\
$M_\mathrm{ejecta}~[\mathrm{M_\odot}]$ & $0.04$ / $0.15$ & $0.03$ / $0.08$ & $0.03$ / $0.15$ & $0.04$ / $0.12$ \\
$E_\mathrm{ejecta}~[10^{50}\,\mathrm{erg}]$ & $3.63$ / $5.56$ & $2.76$ / $5.23$ & $2.84$ / $4.42$ & $3.29$ / $4.98$ \\
$E_\nu~[10^{52}\,\mathrm{erg}]$ & $4.86$/$8.01$ & $5.12$/$9.50$ & $4.75$/$7.74$ & $4.43$/$7.37$ \\
$E_\mathrm{GW}~[10^{46}\,\mathrm{erg}]$ & $1.04$/$2.48$ & $1.25$/$2.96$ & $0.79$/$2.96$ & $0.59$/$1.35$ \\
$T/|W|_\mathrm{pns}^\mathrm{bounce}~[10^{-4}]$ & $6.01$ & $5.63$ & $6.01$ & $5.89$ \\
$L_\mathrm{peak}~[10^{53}\,\mathrm{erg/s}]$ & $6.20$ & $6.34$ & $6.32$ & $6.94$ \\
$\mathrm{M}_{\mathrm{IC}}\,[\mathrm{M}_\odot]$ &0.63 &0.55 &0.63 & 0.61 \\
$\mathrm{R}_{\mathrm{IC}}\,[\mathrm{km}]$ &12.56 & 12.11 & 13.48& 13.03\\
$\mathrm{h}_{\mathrm{feature}}[10^{-20}]$ &1.21 &0.23 &1.17 & 0.55\\
\hline
\end{tabular}
\tablefoot{
$T/|W|_\mathrm{pns}^\mathrm{bounce}$ denotes the ratio between the rotational kinetic energy ($T$) and the gravitational binding energy (expressed as its absolute value, $|W|$) of the PNS evaluated at core bounce. $\mathrm{M}_{\mathrm{IC}}$ and $\mathrm{R}_{\mathrm{IC}}$ denote the mass and radius of the inner core at bounce (see Sect.~\ref{sub:coll}). $\mathrm{h}_{\mathrm{feature}}$ is the rms strain amplitude of the GW signal computed in the 100--300 Hz frequency range and over the time interval 5--50 ms post-bounce (see Sect.~\ref{sub:gw}).
}

\end{table*}

\section{Results}
\label{sec:results}
In this section, we discuss the results of our analysis. Table~\ref{tab:summary_ccsn} provides a summary of the main quantitative results.

\subsection{Collapse and bounce properties}
\label{sub:coll}

We start by analyzing both the temporal evolution of central quantities, in particular $Y_\mathrm{e}$, and radial profiles at bounce as a function of enclosed mass. Unless otherwise stated, time-dependent quantities refer to central values, while profiles are evaluated at the time of bounce.
We first discuss the evolution during collapse, and then characterize the structure of the inner core at bounce.

\begin{figure}
    \centering
    \includegraphics[width=0.49\textwidth]{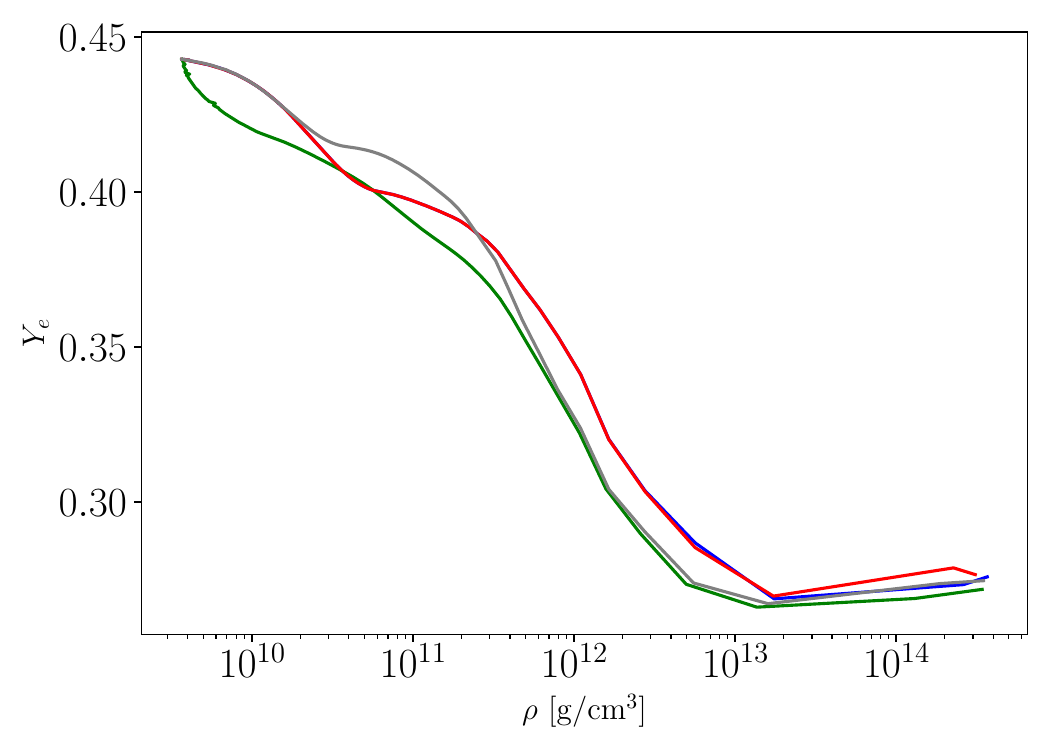}
    \includegraphics[width=0.49\textwidth]{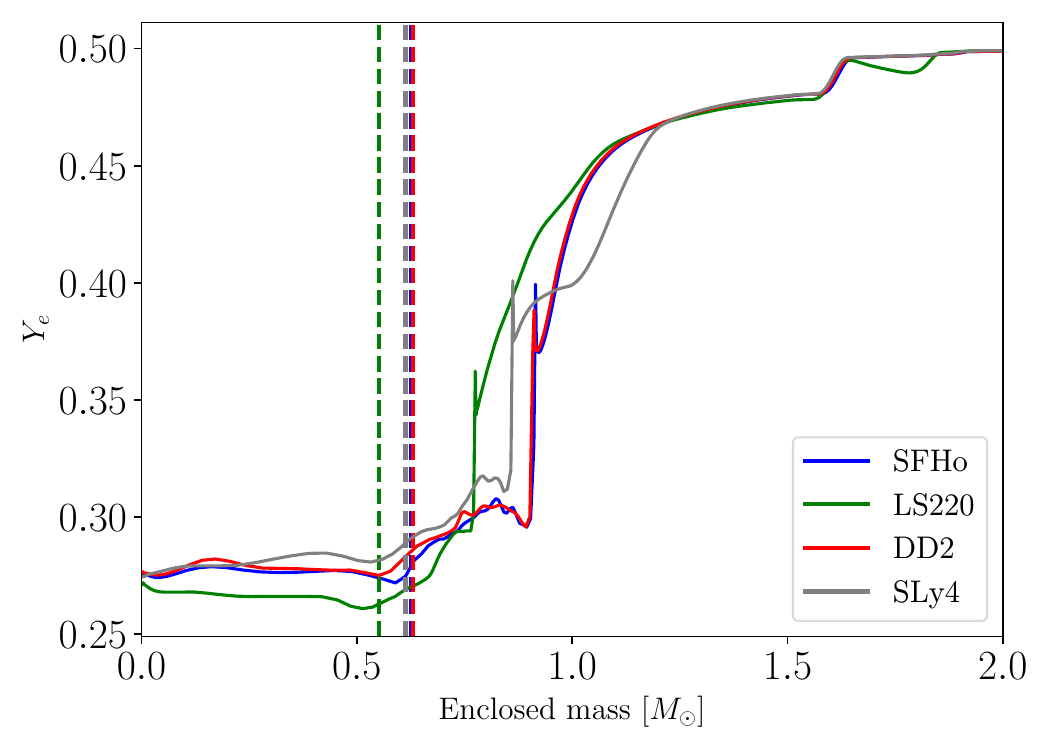}
\caption{Top panel: Evolution of the central electron fraction $Y_e$ as a function of the central baryon density $\rho$ during the collapse phase for the different models. Bottom panel: Electron fraction $Y_e$ as a function of enclosed mass at bounce for the different EoS models. Vertical dashed lines indicate the shock position, defined as the enclosed mass where the entropy first exceeds $s=3$ in the first output following bounce (the SFHo and DD2 lines are nearly coincident). The profiles reflect EoS-dependent differences in deleptonization and inner-core structure.
}
\label{fig:ye_rho}    
\end{figure}

The collapse dynamics is governed by the interplay between thermodynamic evolution and deleptonization. In our models, the most prominent EoS-dependent differences arise from the evolution of the electron fraction $Y_\mathrm{e}$ during collapse. This behavior is illustrated in the top panel of Fig.~\ref{fig:ye_rho}, which shows the central value of $Y_\mathrm{e}$ as a function of the central density.
In particular, the systematically lower $Y_\mathrm{e}$ of LS220 implies a reduced electron degeneracy pressure and a smaller effective Chandrasekhar mass, contributing to the formation of a more compact inner core at bounce. In the following, the inner core is defined as the homologously collapsing, subsonic region enclosed within the sonic point at bounce. SFHo and DD2, which maintain higher $Y_\mathrm{e}$ values throughout collapse, support a more extended homologous core. Although SLy4 exhibits intermediate $Y_\mathrm{e}$ values at low densities and approaches LS220 at $\rho \sim 10^{12}\,\mathrm{g/cm^3}$, differences in the collapse trajectory lead to a reduced integrated deleptonization during the subnuclear phase.

\begin{figure*}[]
    \centering
    \includegraphics[width=1.\linewidth]{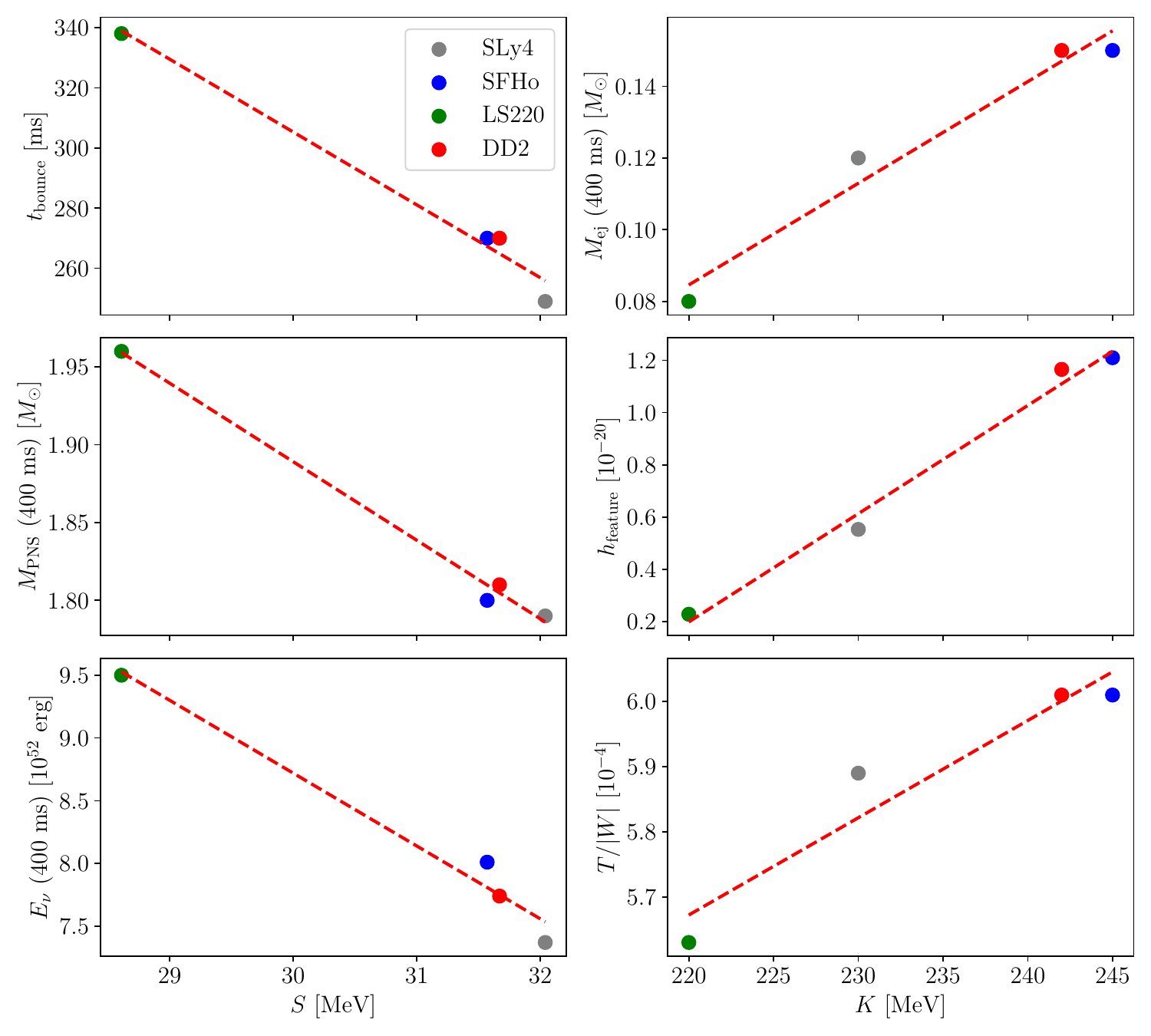}
\caption{Correlations between selected dynamical and observational quantities and the nuclear EoS parameters. The top-left panel shows the correlation between the bounce time $t_{\mathrm{bounce}}$ and the symmetry energy parameter $S$, with a linear fit (red dashed line) characterized by slope $a=-24.201\,\mathrm{ms/MeV}$ and correlation coefficient $R=-0.991$. The middle-left panel shows the correlation between the PNS mass at $t=400\,\mathrm{ms}$ post-bounce and $S$, with slope $a=-0.050\,\mathrm{M_\odot/MeV}$ and correlation coefficient $R=-0.996$. The bottom-left panel shows the correlation between the total emitted neutrino energy at $t=400\,\mathrm{ms}$ post-bounce and $S$, with slope $a=-0.581\,\mathrm{erg/MeV}$ and correlation coefficient $R=-0.987$. The top-right panel shows the correlation between the ejecta mass at $t=400\,\mathrm{ms}$ post-bounce and the incompressibility parameter $K$, with slope $a=0.003\,\mathrm{M_\odot/MeV}$ and correlation coefficient $R=0.983$. The middle-right panel shows the correlation between the averaged gravitational-wave amplitude associated with the prompt-convection signal and $K$, with slope $a=0.041\,\mathrm{MeV}^{-1}$ and correlation coefficient $R=0.994$. The bottom-right panel shows the correlation between the ratio of rotational kinetic energy to gravitational binding energy, $T/|W|$, evaluated at core bounce, and $K$, with slope $a=0.015\,\mathrm{MeV}^{-1}$ and correlation coefficient $R=0.959$. Red dashed lines indicate the corresponding linear fits.}
\label{fig:correlations}
\end{figure*}

The differences in the collapse evolution are reflected in the bounce time, $t_{\mathrm{bounce}}$, defined as the moment when the central density reaches its first maximum (see Table~\ref{tab:summary_ccsn}). The models exhibit variations of several tens of milliseconds, with LS220 collapsing more slowly and SLy4 reaching bounce earlier, while SFHo and DD2 show nearly identical bounce times.
A trend is suggested (top-left panel of Fig.~\ref{fig:correlations}) when $t_{\mathrm{bounce}}$ is considered in combination with the EoS parameter $S$, which characterize the isospin-dependent sector of the nuclear EoS and the neutron-proton asymmetry of nuclear matter. We find an anticorrelation (correlation coefficient $R = -0.991$), although the small number of models considered prevents a statistically robust inference on the slope. Nevertheless, the overall behavior is qualitatively consistent with previous core-collapse studies \citep[e.g.,][]{Fischer2014_EPJA}, which highlight the interplay between deleptonization and $S$ in shaping collapse timescales.
In terms of collapse time, LS220 and SLy4 are the most different cases, which, based on Fig.~\ref{fig:ye_rho} (top panel), may be related to differences in the early $Y_e(\rho_{\mathrm{max}})$ evolution, where most of the collapse time is spent. For $\rho_{\mathrm{max}} \gtrsim 5 \times 10^{11}\,\mathrm{g\,cm^{-3}}$, the curves of the RMF-based models become very similar, as do those of the Skyrme-based models, suggesting that the main differences arise predominantly during the early subnuclear phase of the collapse.

The differences accumulated during collapse directly translate into distinct inner-core configurations at bounce. In order to connect the collapse dynamics discussed in the previous section with the resulting structure at bounce, we analyze the radial profiles as a function of enclosed mass, with particular emphasis on the central electron fraction $Y_\mathrm{e}$, which directly reflects the deleptonization history.
At bounce, the shock forms at systematically different enclosed mass coordinates depending on the EoS. LS220 exhibits shock formation at the smallest enclosed mass, whereas SFHo, DD2, and SLy4 have comparable values, with SLy4 being marginally smaller. However, during the first $\sim1 \,\mathrm{ms}$ post-bounce, the SLy4 shock propagates more rapidly, reaching larger enclosed masses than SFHo and DD2.
The $Y_\mathrm{e}$ profiles shown in Fig.~\ref{fig:ye_rho} (bottom panel) highlight clear EoS-dependent differences. LS220 exhibits the lowest central $Y_\mathrm{e}$, while SLy4, SFHo, and DD2 show progressively higher values. Moving outward in enclosed mass, the profiles reflect the different deleptonization histories accumulated during collapse, with SFHo and DD2 remaining very similar over most of the inner core, and SLy4 approaching intermediate values in the outer regions.
The different shock formation radii also have direct implications for the post-bounce evolution. In particular, since LS220 forms the shock at the smallest enclosed mass, a larger amount of overlying iron-group material is expected to be dissociated as the shock propagates outward, which leads to increased energy losses due to nuclear dissociation.

\subsection{Explosion dynamics}

We first note that all models considered in this work develop explosions at very early post-bounce times. 
The explosion time $t_{\mathrm{exp}}$ is defined as the moment when the polar shock radius reaches $500\,\mathrm{km}$.
This behavior is in marked contrast to the evolution expected for non-rotating or weakly magnetized progenitors of comparable mass, which typically do not exhibit prompt explosions and often rely on prolonged neutrino heating, or may even fail to explode \citep{summaProgenitordependentExplosionDynamics2016,glasThreedimensionalCorecollapseSupernova2019}.
In our simulations, the early onset of the explosion can be directly attributed to the presence of rapid rotation and strong magnetic fields, which enable efficient magnetorotational mechanism to tap the available rotational energy and launch bipolar outflows. This outcome is fully consistent with the scope of the present study, which is specifically aimed at exploring the dynamics and multi-messenger signatures of magnetorotational CCSNe, where early MHD-driven explosions constitute a defining feature.

\begin{figure}[]
    \centering
    \includegraphics[width=0.49\textwidth]{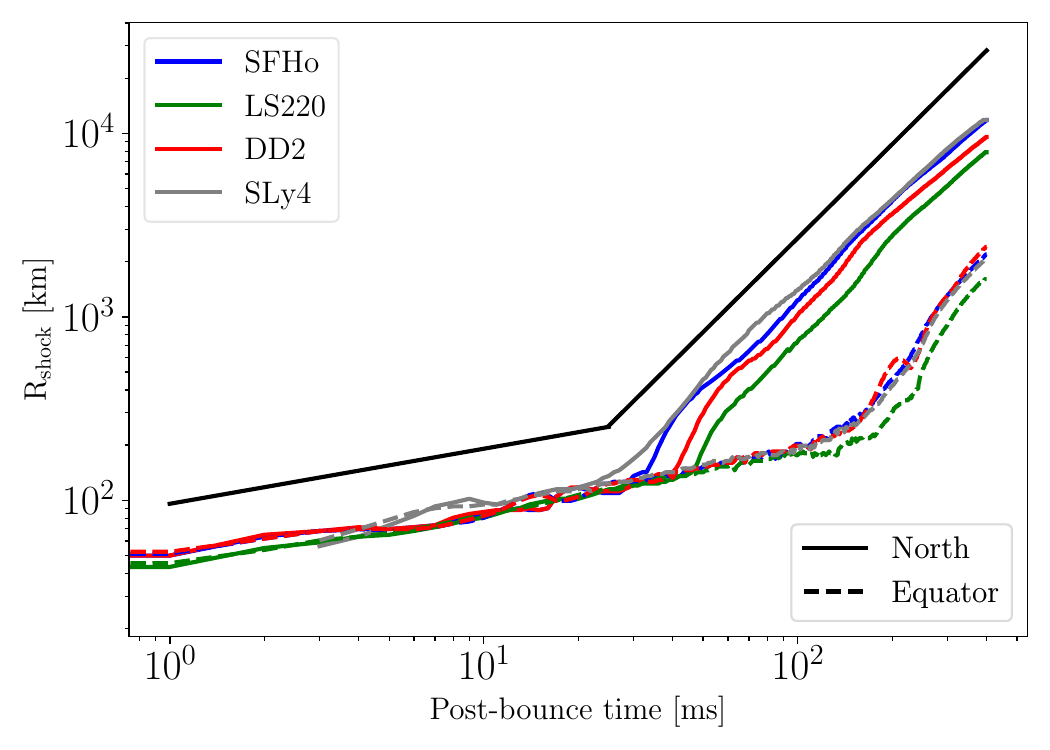}
\caption{Time evolution of the shock radius at the north pole (solid lines) and at the equator (dashed lines) for the four EoS models. Black solid lines show reference power-law scalings of the shock expansion with slopes of 0.3 and 1.7.}
    \label{fig:radii_EoS}
\end{figure}

In this context of early RMHD-driven explosions, we analyze the influence of the nuclear EoS on the shock evolution, focusing on the shock radius along the north polar direction and the equatorial plane (see Fig.~\ref{fig:radii_EoS}, solid and dashed lines, respectively).
Among the models considered, SLy4 exhibits the fastest shock expansion, while LS220 shows the slowest evolution. As discussed in Sec.~\ref{sub:coll}, this behavior is rooted in the different inner-core structure at bounce. In particular, the location of shock formation at smaller enclosed mass in LS220 leads to increased dissociation of overlying material, which delays the onset of efficient shock expansion.
Beyond the EoS dependence, the shock evolution follows a clear piecewise power-law behavior in time. During the stalling phase, before shock revival, the evolution is well described by a shallow power law with slope $\sim 0.3$, while after revival, the expansion steepens significantly, approaching a slope of $\sim 1.7$. These scalings are shown as reference fits in Fig.~\ref{fig:radii_EoS}.
The comparison between polar and equatorial directions highlights a strong asphericity of the explosion. The shock expansion is systematically faster along the polar axis, consistent with a more efficient MHD-driven acceleration in the polar funnel. In contrast, the equatorial region exhibits a significantly delayed and more gradual revival, reflecting the anisotropic nature of the MHD mechanism.

\begin{figure}[]
        \centering
        \includegraphics[width=0.49\textwidth]{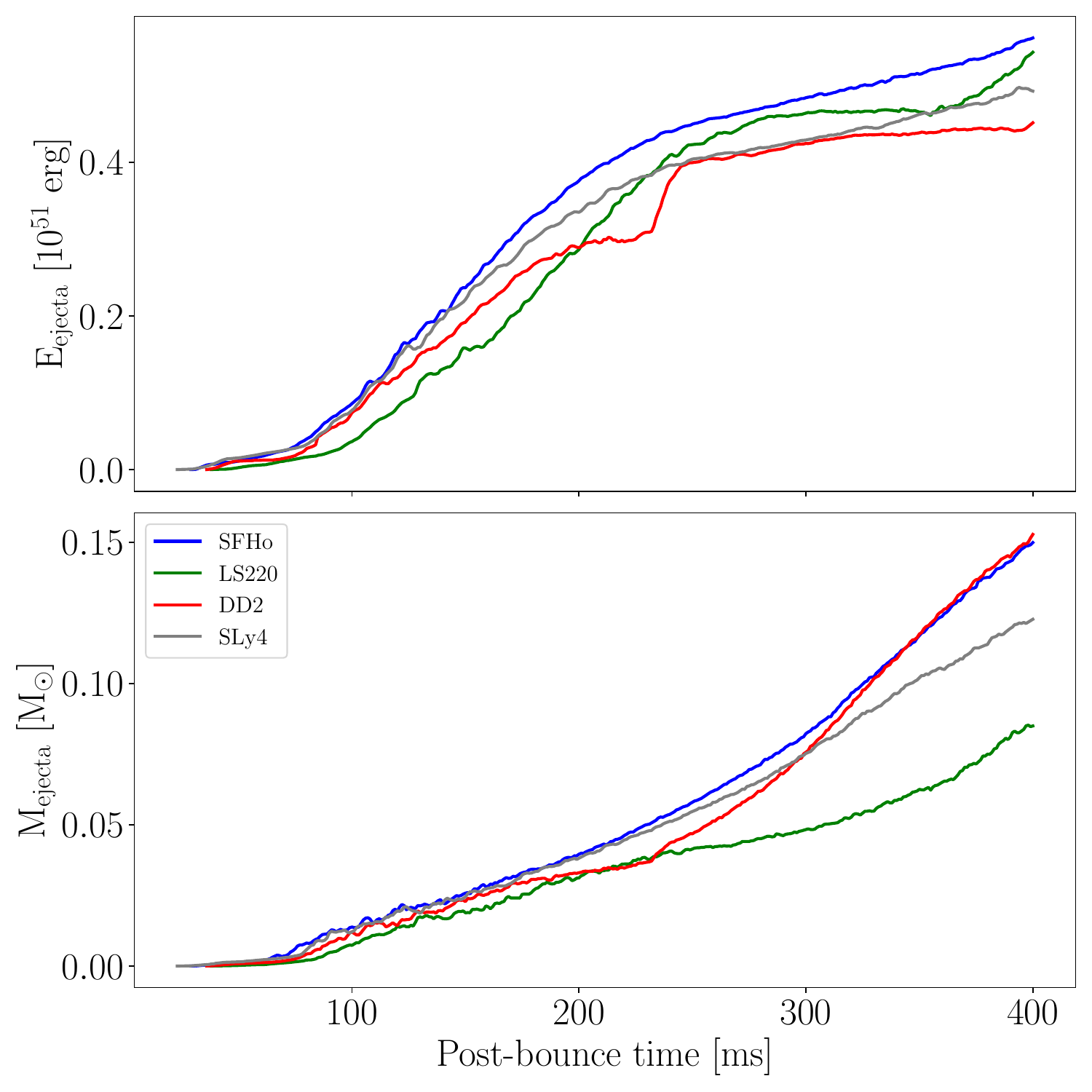}
            \caption{\small Time evolution of the total energy of the ejecta (top) and of the total mass of unbound ejecta (bottom) for the four EoS models. 
            }
            \label{fig:ejecta_e_EoS}
\end{figure}

Figure~\ref{fig:ejecta_e_EoS} shows the total energy (upper panel) and mass (lower panel) of the gravitationally unbound ejecta for the different EoS models. Ejecta are identified as fluid elements with positive total energy density and positive radial velocity. We observe that the most massive and energetic ejecta are produced by the softer EoS, with SFHo in particular yielding the highest values in both quantities.

The differences between the models are ultimately linked to the EoS-dependent inner-core properties at bounce, which set the initial conditions for the subsequent magneto-hydrodynamical evolution.
Of particular importance is the post-bounce compactness of the PNS. In the case of SFHo, the stronger contraction leads to a more efficient amplification of the toroidal magnetic field through differential rotational winding (the $\Omega$-mechanism), as well as of non-radial field components through flux freezing. The resulting magnetic stresses enhance angular-momentum transport and facilitate a more efficient extraction of rotational energy, ultimately powering a more energetic outflow.
SLy4, in contrast, forms a less compact PNS, leading to weaker rotational winding during the post-bounce evolution and therefore to a less efficient amplification of the toroidal magnetic field. This results in less efficient rotational-energy extraction and correspondingly lower ejecta energies and masses.
LS220 produces the least energetic and least massive ejecta at early times. This behavior is linked to its delayed core dynamics and deeper shock formation, which may reduce the efficiency of energy transfer to the outflow and prolong the accretion phase onto the PNS. Although simultaneous accretion and outflow are present in all models during the early post-bounce evolution, this phase appears to persist longer in the LS220 case.
DD2 also exhibits a transient feature around $250\,\mathrm{ms}$ post-bounce, where a fraction of equatorial material becomes unbound, leading to a sudden increase in ejecta mass and energy. This highlights the role of local dynamical effects in modulating the otherwise global EoS-dependent trends.
Finally, we find a strong correlation between the ejecta mass at $t=400\,\mathrm{ms}$ and the EoS parameter $K$ (top-right panel of Fig.~\ref{fig:correlations}), with a correlation coefficient $R = 0.983$.
This suggests that the long-term ejecta production is tightly linked to the underlying stiffness properties of the EoS, which regulate both PNS compactness and the efficiency of rotational-energy extraction.

\subsection{PNS properties}
\begin{figure*}
    \centering
    \includegraphics[width=1.\linewidth]{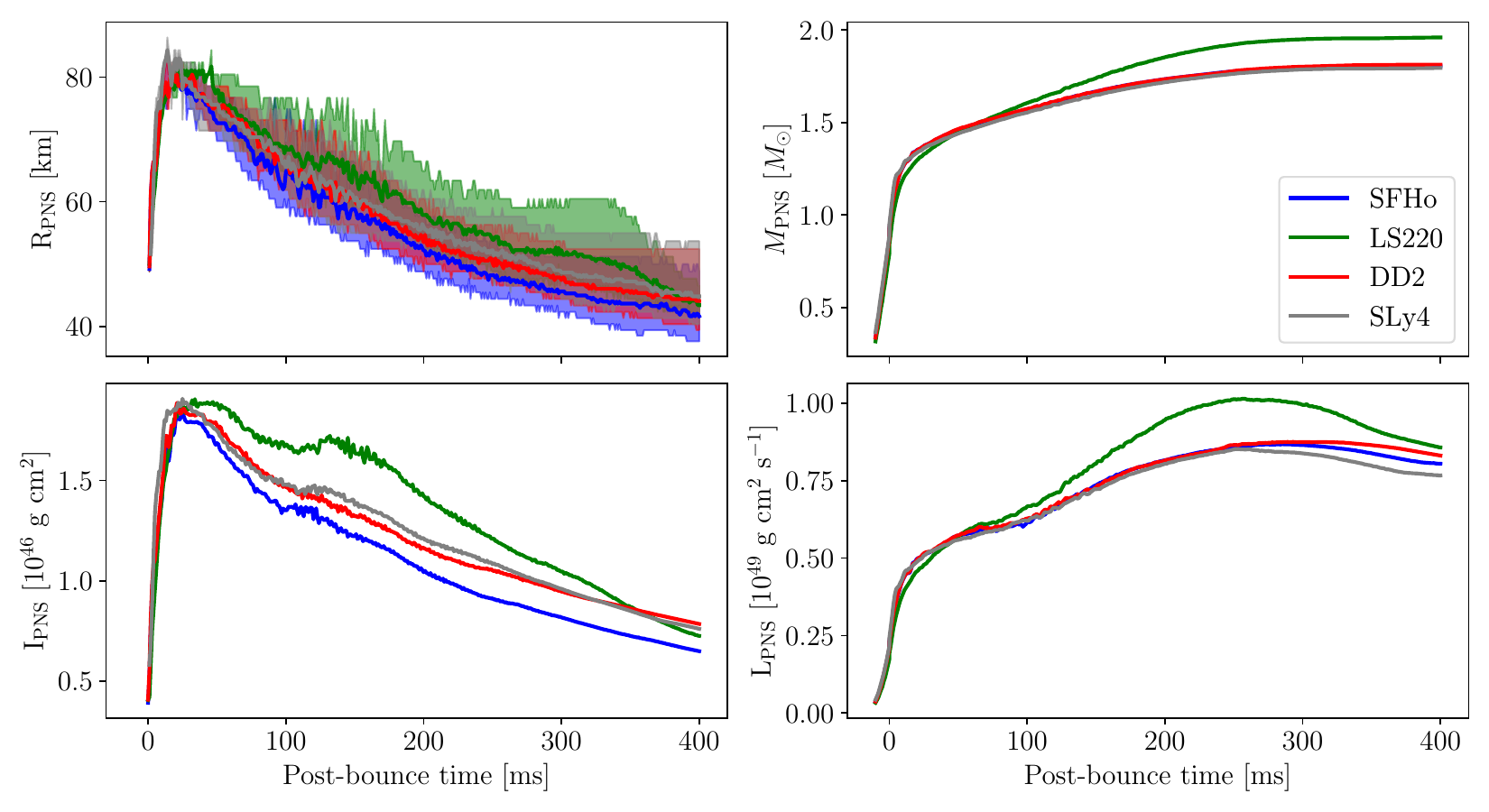}

    \caption{Time evolution of various PNS properties for the four EoS models. Top-left panel: the mean radius (solid lines) and the minimum/maximum range as the shaded area for the four EoS models. Top-right: the total mass contained in the PNS volume. Bottom-left: the total moment of inertia with respect to the polar axis. Bottom-right: the total angular momentum of the PNS. 
 }
    \label{fig:pns prop}
\end{figure*}

In the present subsection, we will discuss the main PNS properties summarized in Fig.~\ref{fig:pns prop}.

The PNS radius, defined by the isodensity contour at $10^{11}\,\mathrm{g\,cm^{-3}}$, is shown in the top-left panel. The mean radius, computed as the arithmetic average over the northern polar axis, southern polar axis, and equatorial plane, is indicated by the solid curve. The shaded region spans the range between the radius at the north pole (minimum) and at the equator (maximum), thereby illustrating the degree of PNS asphericity.
Among the models, LS220 produces the least compact PNS, with systematically larger radii and a more extended polar-to-equator structure. In contrast, SFHo forms the most compact configuration, while DD2 and SLy4 exhibit intermediate behavior. The evolution of the radius reflects the combined effects of mass accretion, rotation, magnetic stresses, and neutrino cooling, with the cumulative neutrino emission (see Table~\ref{tab:summary_ccsn}) reflecting the overall cooling history that contributes to PNS contraction.

These structural differences are directly reflected in the evolution of the moment of inertia and angular momentum (bottom-left and bottom-right panels). SFHo, being more compact, exhibits the lowest moment of inertia, while LS220 maintains the largest values over most of the evolution, consistently with its more extended PNS structure. SLy4 and DD2 remain comparable. Although LS220 starts from the largest moment of inertia and angular momentum, it also exhibits the steepest decline during the post-bounce evolution. Around $\sim300$--$350\,\mathrm{ms}$ post-bounce, the evolution shows a more pronounced decrease in both $L_{\mathrm{PNS}}$ and the moment of inertia, coincident with a contraction of the PNS radius and an increase in ejecta energy. This behavior suggests a phase of enhanced angular-momentum redistribution and extraction, possibly associated with more efficient magnetic braking as the PNS becomes more compact. For SFHo, DD2, and SLy4 the evolution of $L_{\mathrm{PNS}}$ is instead broadly similar up to $\sim250\,\mathrm{ms}$ post-bounce, after which SLy4 shows a slightly more pronounced decline in angular momentum, while SFHo remains intermediate and DD2 retains marginally higher angular momentum at late times.

The time evolution of the PNS mass (top-right panel) provides complementary information on the long-term evolution of the system. LS220 exhibits a systematically higher mass accretion, leading to a more massive PNS compared to the other models, which remain within a narrower range. This behavior is consistent with its delayed collapse evolution and weaker early explosion dynamics, which prolong the accretion phase and reduce the efficiency with which infalling material is redirected into the outflow.
From Appendix~\ref{App:m-r}, which shows the maximum gravitational mass supported by each EoS, we note that SLy4, SFHo, and LS220 have similar maximum cold neutron star masses. The larger PNS mass reached in LS220 therefore suggests a comparatively smaller margin with respect to the maximum supported mass, although longer simulations would be required to assess whether collapse to a black hole eventually occurs. In contrast, DD2 yields final PNS masses comparable to those of SFHo and SLy4, but its significantly higher maximum supported mass implies a larger stability margin under similar conditions.
A further insight into the long-term evolution of the system can be obtained by examining the interplay between the PNS mass and the underlying microphysics. In particular, the role of isospin-dependent effects can be illustrated by correlating the PNS mass at $t=400\,\mathrm{ms}$ with the parameter $S$ (mid-left panel of Fig.~\ref{fig:correlations}). A negative correlation is obtained, with a Pearson coefficient $R=-0.996$, although this result should be interpreted with caution given the limited sample size.
This trend suggests that models characterized by larger $S$ tend to produce less massive PNS at late times. Within the present set of simulations, this behavior appears to reflect differences in the efficiency of mass accumulation during the post-bounce accretion phase. 

The structural differences discussed above are also reflected in the rotational support of the PNS at core bounce. As shown in the bottom-right panel of Fig.~\ref{fig:correlations}, the ratio $T/|W|$ exhibits a positive correlation with the EoS incompressibility parameter $K$, with a Pearson correlation coefficient of $R=0.959$. Stiffer equations of state therefore retain a larger fraction of rotational support at bounce.
This behavior is consistent with the larger PNS radii and moments of inertia obtained for stiffer EoSs. Their lower compactness reduces the efficiency of angular-momentum redistribution during collapse, allowing a larger fraction of the initial rotational energy to remain stored in ordered rotation rather than being transferred to local turbulent motions or magnetic fields. The resulting rotational state provides the initial conditions for the subsequent post-bounce evolution of the magnetic and kinetic energy reservoirs discussed below.
Although our simulations are axisymmetric, the inferred differences in $T/|W|$ suggest that, in fully three-dimensional models, stiffer EoSs could evolve closer to the threshold for non-axisymmetric rotational instabilities, with possible consequences for angular-momentum transport and gravitational-wave emission.

\begin{figure}
        \centering
\includegraphics[width=0.49\textwidth,clip,trim={0 1.7cm 0 0}]{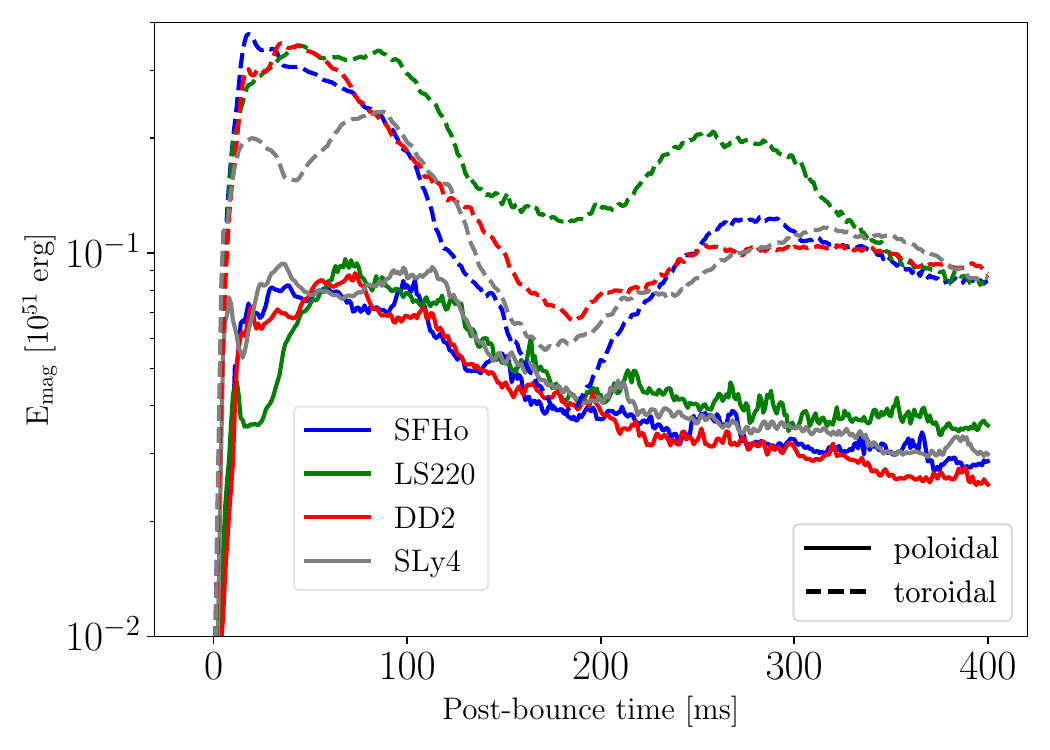}
        \includegraphics[width=0.49\textwidth]{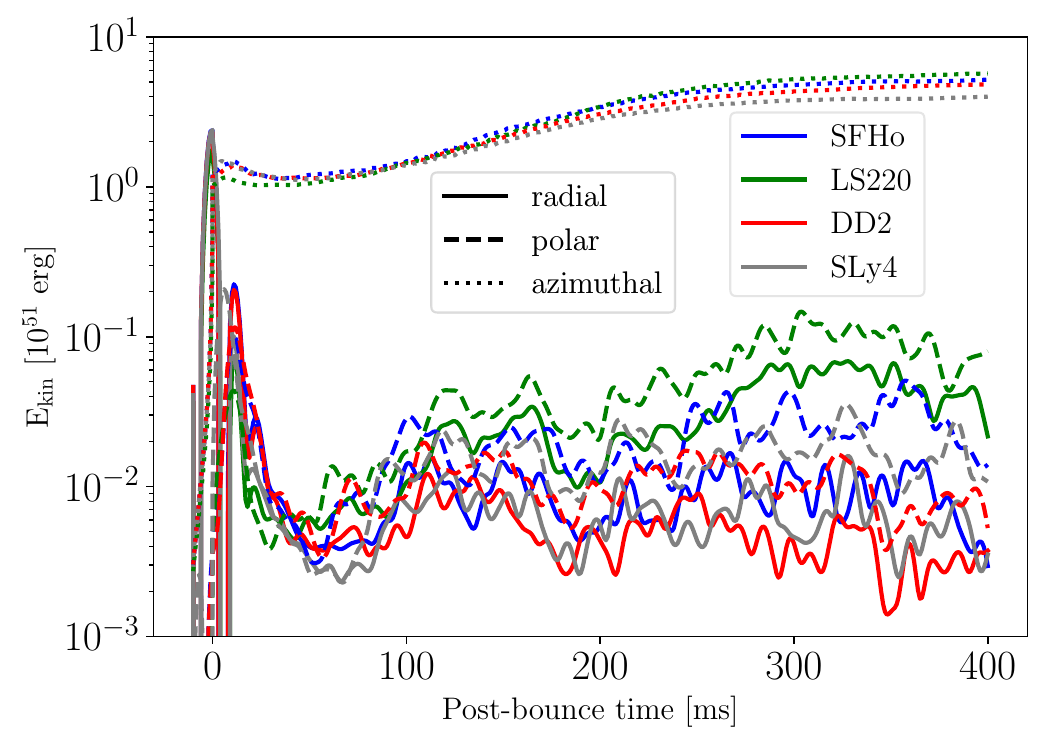}
        
 \caption{\small Top: Time evolution of the magnetic energy within the PNS for the four EoS models. Bottom: Time evolution of the kinetic energy within the PNS for the four EoS models. A low-pass Butterworth filter has been applied to the radial and polar components of the kinetic energy to suppress numerical noise and improve the readability of the long-term evolution.}
     \label{fig:pns_kin_EoS}
\end{figure}

To examine how this rotational energy is subsequently redistributed within the PNS, we now turn to the evolution of the magnetic and kinetic energy components.
All magnetic and kinetic (Fig.~\ref{fig:pns_kin_EoS}, upper and lower panels, respectively) energies are computed via volume integration over the PNS.
The magnetic energy is decomposed into poloidal and toroidal components:
\begin{equation}
E_{\mathrm{mag,pol}} = \int_{V_{\mathrm{PNS}}} \frac{B_r^2 + B_\theta^2}{8\pi} \, dV,
\qquad
E_{\mathrm{mag,tor}} = \int_{V_{\mathrm{PNS}}} \frac{B_\phi^2}{8\pi} \, dV.
\label{eq:emag}
\end{equation}

The kinetic energy components are defined as
\begin{equation}
E_{\mathrm{kin},i} = \int_{V_{\mathrm{PNS}}} \frac{1}{2} \rho v_i^2 \, dV,
\label{eq:ekin}
\end{equation}
with $i \in \{r,\theta,\phi\}$, as relativistic corrections remain small ($\gamma \approx 1$) throughout our simulations.

The $\phi$ component of the magnetic energy in the PNS (Fig.~\ref{fig:pns_kin_EoS}, upper panel) exhibits a similar evolution across all models. A common decrease is observed around $\sim100\,\mathrm{ms}$ post-bounce, possibly reflecting enhanced angular-momentum redistribution and energy transfer during this phase. This feature is present in all cases but is most pronounced in SFHo, indicating a stronger reduction of the toroidal magnetic field component during this phase. In contrast, LS220, while exhibiting the largest toroidal magnetic energy, shows the same transient decrease but maintains systematically higher absolute values, suggesting that a larger fraction of the magnetic energy remains stored in the azimuthal component within the PNS. 
Importantly, these variations in the internal magnetic energy partitioning do not translate into significant changes in the global angular momentum evolution, which remains broadly similar among SFHo, DD2, and SLy4. However, changes in the moment of inertia indicate a concurrent structural readjustment of the PNS, suggesting that internal RMHD reconfigurations are accompanied by modifications of the stellar structure without substantially altering the total angular momentum budget. No clear corresponding feature is observed in the global explosion energy within the time window considered, indicating that this phase does not have an immediate impact on the large-scale outflow energetics.

The post-bounce magnetic field configuration is dominated by a strong toroidal component amplified via differential rotation ($\Omega$-winding) in the PNS. In all models, magnetic energy is therefore primarily stored in large-scale toroidal structures concentrated in the outer PNS layers, while the poloidal component remains subdominant and dynamically less relevant for the global energy budget.
The rotation profile is strongly differential in all cases. The angular velocity increases from the inner core toward the outer layers of the PNS, reaching a maximum close to the PNS surface, where values exceeding $\sim10^{3}\,\mathrm{rad\,s^{-1}}$ are attained in all models. Outside the PNS boundary, $\Omega$ drops sharply, clearly separating the rapidly rotating core from the more slowly rotating accretion region. While this qualitative behavior is common to all EoSs, quantitative differences in the peak values of $\Omega$ reflect the different compactness and moment of inertia of the PNS.
This differential rotation is the main driver of magnetic-field amplification via the $\Omega$-winding mechanism and sets the efficiency of angular momentum transport through magnetic stresses. The PNS surface thus acts as the main region where rotational energy is converted into magnetic energy and subsequently redistributed within the system.

To conclude this analysis, we examine the radial and polar components of the kinetic energy, shown in Fig.~\ref{fig:pns_kin_EoS}, which provide a useful tracer of convective motions within the PNS. 
At earlier times ($t \lesssim 10$--$20\,\mathrm{ms}$ post-bounce), all models exhibit a phase of prompt convection driven by negative entropy gradients behind the shock. In this phase, LS220 shows comparatively weaker convective kinetic energy than the other EoSs, indicating a less pronounced development of prompt convective activity.
At later times ($t \gtrsim 80\,\mathrm{ms}$ post-bounce), the system transitions to Ledoux convection within the PNS. A clear enhancement of convective activity is observed in the LS220 model. This behavior can be primarily attributed to its larger PNS volume, which leads to a more extended region satisfying the Ledoux instability criterion and therefore supports stronger sustained convective motions.
Among the other three EoSs, SLy4 and SFHo exhibit similar levels of convective kinetic energy, while DD2 shows slightly lower values. These differences are likely related to variations in the entropy and lepton-fraction gradients, which regulate the onset and strength of Ledoux convection.
Further details on the spatial structure of the entropy and electron-fraction gradients during the prompt convection phase ($t \sim 10\,\mathrm{ms}$ post-bounce) are provided in Appendix~\ref{App:BV}, where we also show the corresponding Brunt–Väisälä stability indicator.

\subsection{Neutrino emission}
We now turn to the analysis of the neutrino emission, which reflects the interplay between the thermodynamic state of the post-bounce matter, the evolving opacity structure, and the multidimensional accretion flow.

\begin{figure}
        \centering
        \includegraphics[width=0.49\textwidth]{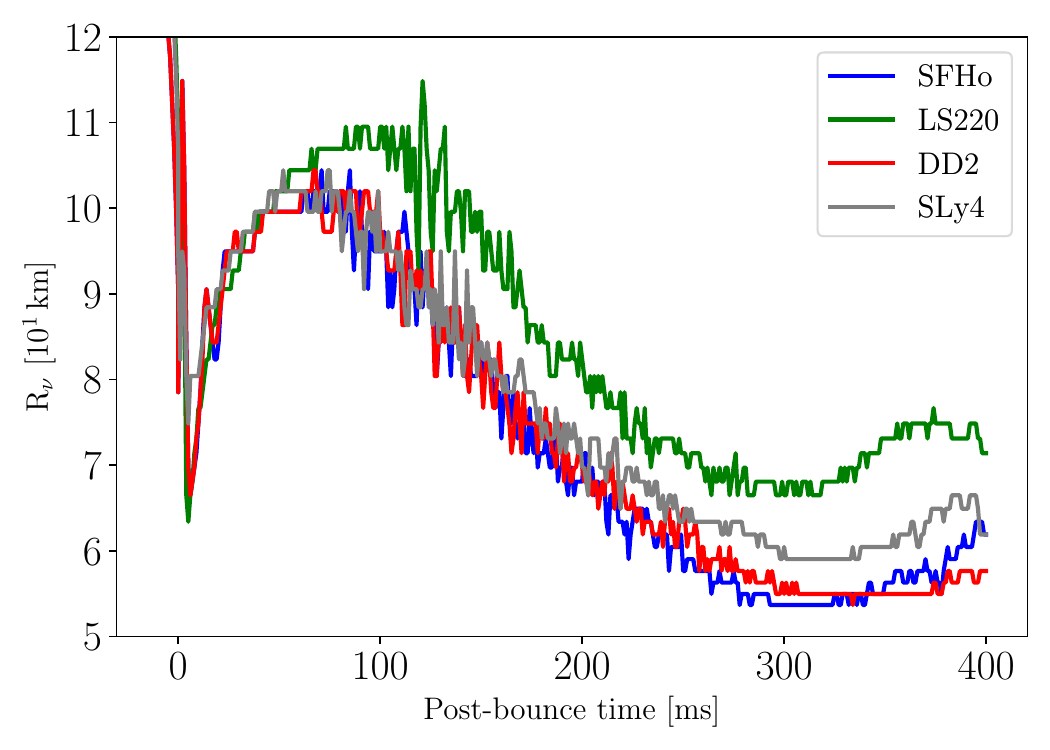}
    \caption{\small Time evolution of the electron-neutrinosphere radius at the equator for the four EoS models.  
    }
     \label{fig:neu_radii_EoS}
\end{figure}

The neutrino emission is closely connected to the spatial extension of the semi-transparent region, as characterized by the neutrinosphere, defined as the radius at which the optical depth reaches $\tau = 2/3$ (Fig.~\ref{fig:neu_radii_EoS}). 
The optical depth is computed by radially integrating an energy-averaged opacity, where the averaging is performed over the neutrino energy spectrum using the neutrino energy density in each energy bin as a weight. 
The resulting neutrinosphere is determined for each neutrino species and angular direction and corresponds to a spectrum-averaged surface rather than an energy-dependent one. 
In Fig.~\ref{fig:neu_radii_EoS} we focus on the electron-neutrino neutrinosphere evaluated along the equatorial direction, as it is representative of the overall ordering among models, with the other neutrino species showing qualitatively similar trends. In this sense, LS220 exhibits systematically larger equatorial radii compared to the other models, which may indicate a more extended region of last scattering and a potentially larger effective emitting region. Among the remaining models, SLy4 also shows slightly larger equatorial radii than DD2 and SFHo.

All models exhibit a phase of increasing neutrinosphere radius starting at approximately $300\,\mathrm{ms}$ post-bounce. 
Although the PNS is globally contracting, the neutrinosphere is determined by the optical depth rather than by the matter radius alone. 
In our rapidly rotating models, continued equatorial accretion after the onset of the explosion sustains a layer of relatively opaque material in the outer regions, which shifts the $\tau = 2/3$ surface to larger radii. 
This indicates that the neutrinosphere evolution is influenced not only by global contraction, but also by anisotropic accretion during the post-bounce phase.

\begin{figure}[b]
    \centering
    \includegraphics[width=\linewidth]{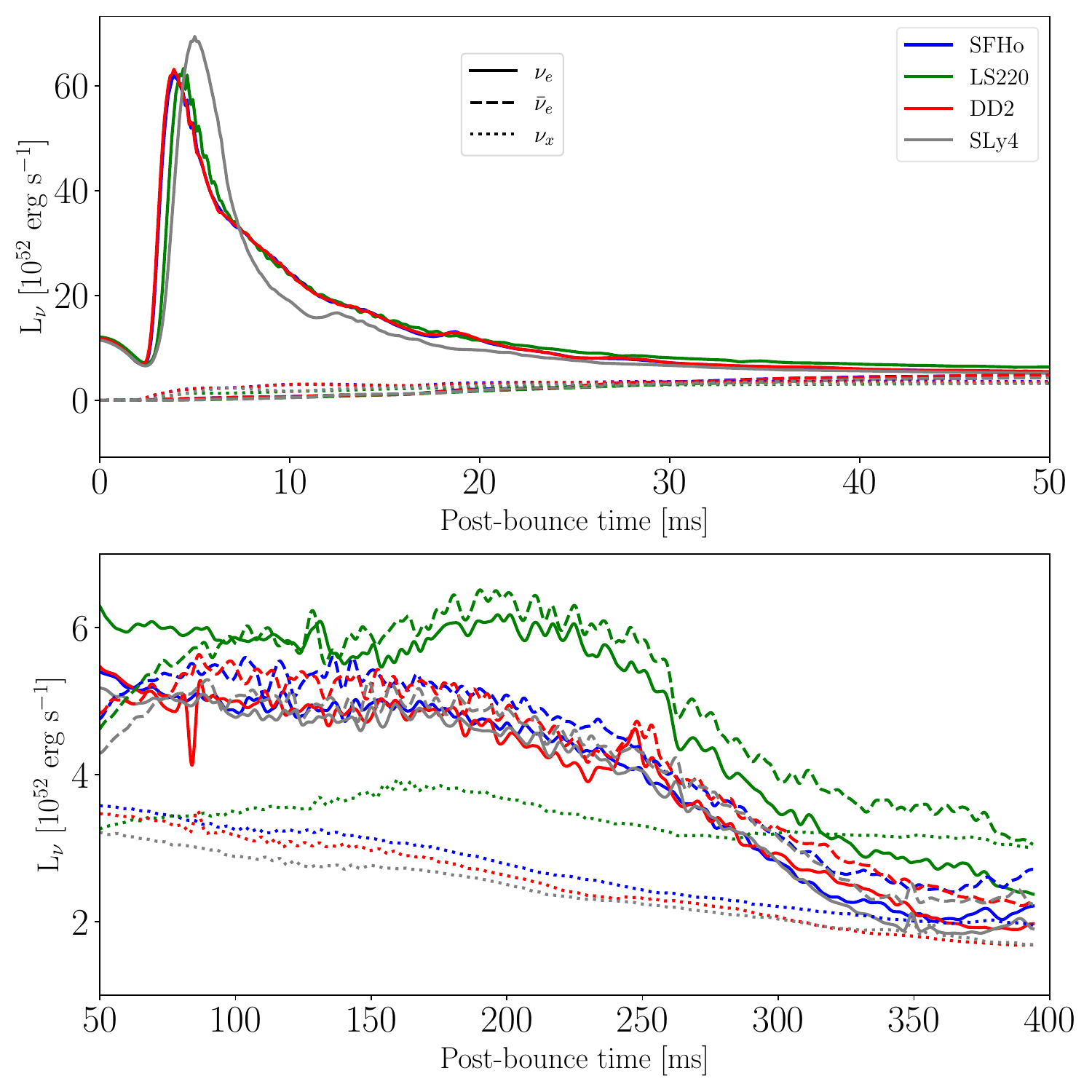}
\caption{\small Top: Time evolution of the average neutrino luminosity up to $50\,\mathrm{ms}$ p.b.
Bottom: Time evolution of the average neutrino luminosity from $50\,\mathrm{ms}$ p.b. LS220 shows the highest emission of each species. A low-pass Butterworth filter has been applied to the luminosity curves to suppress numerical noise and improve the readability of the long-term evolution. The sharp, short-lived spikes observed in the unfiltered luminosity evolution are attributed to numerical noise in the extraction procedure and are not considered physically meaningful.
}
    \label{fig:lum_eq_EoS}
\end{figure}

We now turn to the analysis of neutrino emission, computed at a radius of $500\,\mathrm{km}$ in the equatorial plane, which reflects the interplay between the thermodynamic state of the post-bounce matter, the evolving opacity structure, and multidimensional accretion flows.

The most striking feature is the height of the breakout burst peak, which is noticeably higher in the SLy4 model (upper panel in Fig.~\ref{fig:lum_eq_EoS}).
To better understand these differences, we note that the central density at bounce follows the ordering 
$\rho_c^{\mathrm{bounce}}(\mathrm{LS220}) > \rho_c^{\mathrm{bounce}}(\mathrm{SLy4}) > \rho_c^{\mathrm{bounce}}(\mathrm{DD2}) \approx \rho_c^{\mathrm{bounce}}(\mathrm{SFHo})$, 
with relatively small differences between SLy4, DD2 and SFHo. 
Since deleptonization during collapse is tightly coupled to the compression history of the core, higher central densities lead to a larger trapped electron fraction and neutrino content at bounce, which sets the available energy scale for the breakout emission.
In addition, the larger neutrinosphere radius observed in SLy4 shortly after bounce may further increase the effective emitting area, potentially contributing to its higher neutrino luminosity under the assumption of comparable neutrinosphere temperatures across models.
Another possible contribution is related to the evolution of the electron fraction. At bounce, SLy4 exhibits a slightly higher $Y_e$ than the other models (Fig.~\ref{fig:ye_rho}), suggesting a larger reservoir of electrons available for subsequent capture. During the first $\sim10\,\mathrm{ms}$ post-bounce, however, $Y_e$ decreases more markedly and becomes lower than in the other models in the region between $\sim0.6$ and $0.8\,M_\odot$ (Fig.~\ref{fig:s_ye_mass}). This behavior is consistent with enhanced electron captures in these layers, and hence with a larger release of electron neutrinos contributing to the breakout burst. In addition, a more rapid propagation of the shock through the semi-transparent region may further sharpen the neutrino release, leading to a higher peak luminosity. This faster shock evolution may also affect the dissociation of iron-group nuclei and the subsequent transition of the shocked material toward the semi-transparent regime, potentially contributing to a more abrupt release of the trapped electron neutrinos. At present, these interpretations remain qualitative, and a dedicated investigation including a broader set of EoSs will be required to assess their relative importance.

\begin{figure}
    \centering
    \includegraphics[width=\linewidth]{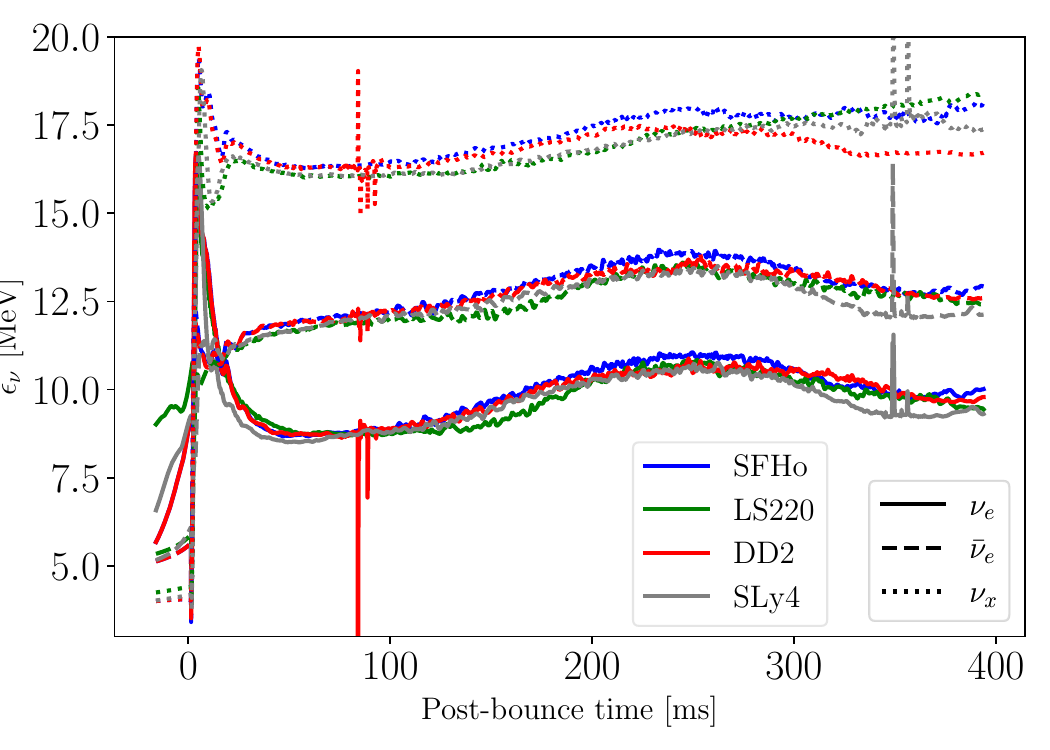}
    \caption{\small: Time evolution of the averaged RMS neutrino energy for the four EoS models and for the different species, indicating the temperature of the layer emitting the neutrinos. 
A few short-lived spikes in the RMS neutrino energy are likely related to numerical noise and do not correspond to physical features of the signal.
}
    \label{fig:rms}
\end{figure}

At later times, the neutrino luminosity reflects the combined effects of the emitting surface and the local neutrino flux, with
$L_\nu = 4\pi R_\nu^2 F_\nu$,
where both $R_\nu$ and $F_\nu$ vary across models. 
The flux is primarily set by the local thermodynamic state and composition of the semi-transparent region and is therefore sensitive to both the EoS and the multidimensional flow dynamics.
An additional and persistent contribution arises from accretion-powered emission, which remains active as long as mass continues to be funneled onto the PNS. 
In our rapidly rotating models, accretion is strongly anisotropic and persists preferentially in the equatorial plane even after the onset of shock revival, sustaining a non-negligible luminosity at late times.
Within this framework, LS220 systematically exhibits the highest neutrino luminosity in the post-explosion phase (lower panel in Fig.~\ref{fig:lum_eq_EoS}). 
This behavior results from the combination of a larger effective emitting surface and a sustained accretion rate, which together enhance both the PNS cooling and accretion-driven contributions to the emission.
The RMS neutrino energies (Fig.~\ref{fig:rms}) provide a complementary diagnostic of the thermodynamic conditions in the emitting region, indicating that lower RMS energies correspond to cooler effective decoupling layers. 
In the case of LS220, the higher luminosity is therefore not driven by higher temperatures but rather by the larger emitting area and the continued accretion activity.

To further quantify the global emission, we consider the total emitted neutrino energy $E_\nu$ reported in Table~\ref{tab:summary_ccsn} \citep[for this quantity, see][]{Nagakura2021}.
We find a clear hierarchy, with LS220 producing the largest total emission, followed by SFHo and DD2, and finally SLy4. 
A strong anticorrelation with $S$ is observed (bottom-left panel of Fig.~\ref{fig:correlations}), with a Pearson coefficient $R = -0.987$. 
This trend can be understood as the combined result of collapse and post-bounce evolution: models with smaller $S$ undergo more efficient deleptonization, leading to a larger trapped lepton reservoir at bounce and sustaining a stronger neutrino emission during both cooling and accretion phases.

\subsection{Gravitational wave signal}
\label{sub:gw}
\begin{figure}[b]
    \centering
    \includegraphics[width=0.49\textwidth]{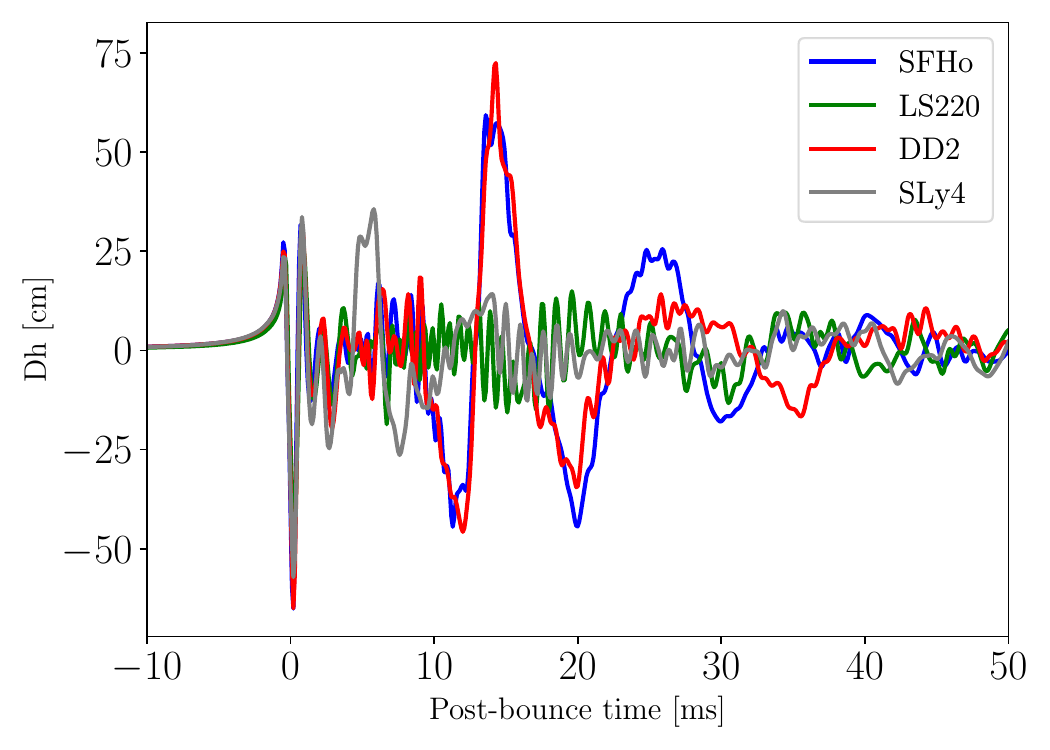}
\caption{Gravitational wave signals shortly after core bounce for the four EoS models. 
}
    \label{fig:gw_short_EoS}
\end{figure}

Throughout this work, we show the distance-rescaled strain $Dh$, in order to remove the trivial geometrical dependence on the source distance. 
Details on the GW extraction procedure are given in Appendix~\ref{App:GW}.

Regarding the GW signal, the emission associated with core bounce is similar in overall shape across all four EoS models (Fig.~\ref{fig:gw_short_EoS}), but the depth of the bounce minimum differs noticeably between them. In particular, LS220 produces the least intense signal, consistent with its less compact and more extended PNS at bounce. The larger PNS radius is expected to reduce the characteristic dynamical accelerations during core bounce, leading to a weaker variation of the mass quadrupole and hence to a shallower GW minimum.

Of particular interest here is the immediate ($\sim50\,\mathrm{ms}$) post-bounce GW signal, which highlights the different response of each EoS to the same dynamical perturbation. 
In the case of LS220, we observe a strong, high-frequency, and time-modulated GW emission, originating from bounce-induced oscillation modes of the PNS. 
In contrast, SLy4 exhibits a brief low-frequency emission phase immediately after bounce, likely associated with prompt convection. 
The prompt-convection signal in SLy4 has a significantly shorter duration than in SFHo and DD2, and is punctuated by high-frequency peaks similar to those observed in LS220, which are associated with oscillation modes of the PNS.
SFHo and DD2 once again exhibit more similar behavior. Prompt convection excites low-frequency modes between approximately $20$ and $50\,\mathrm{ms}$ post-bounce, which are more rapidly damped in DD2. In contrast, SFHo sustains these modes for a longer duration, likely due to the matter's ability to oscillate coherently over longer timescales. This coherence may be favored by a softer EoS, which allows the system to compress to a higher extent, thus allowing for oscillation modes with longer wavelengths.

\begin{figure*}[]
    \centering
    \begin{minipage}[t]{0.45\textwidth}
        \centering
        \includegraphics[width=1.\linewidth,clip,trim={0 2.0cm 0 0}]{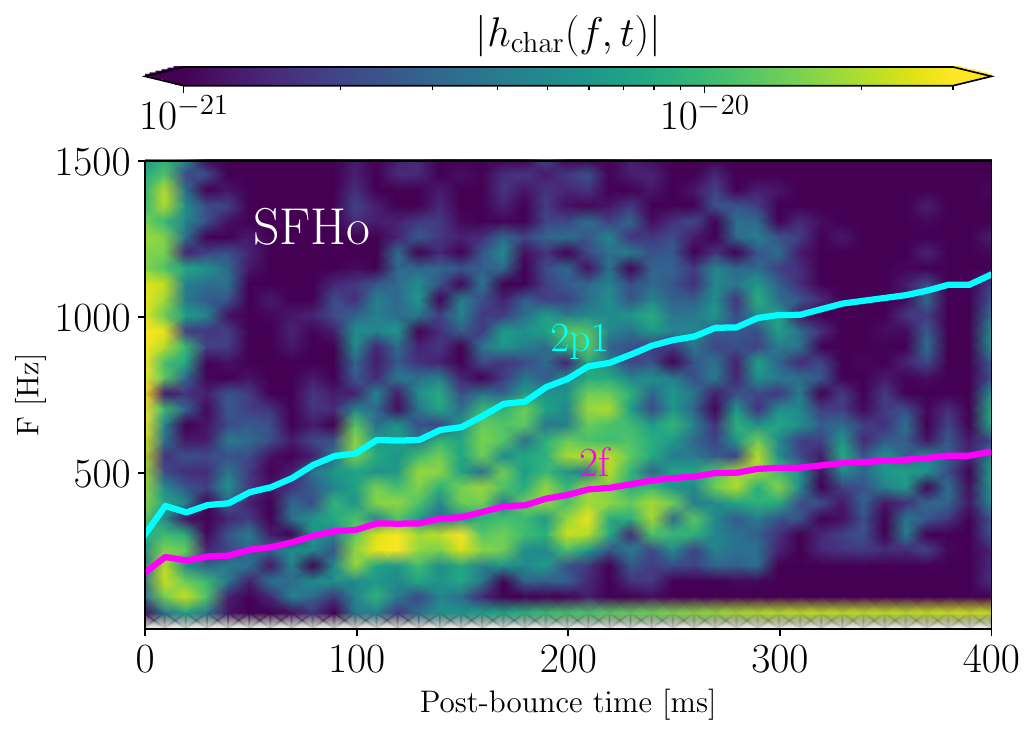}
        \includegraphics[width=0.97\linewidth,clip,trim={0 1.66cm 0 0}]{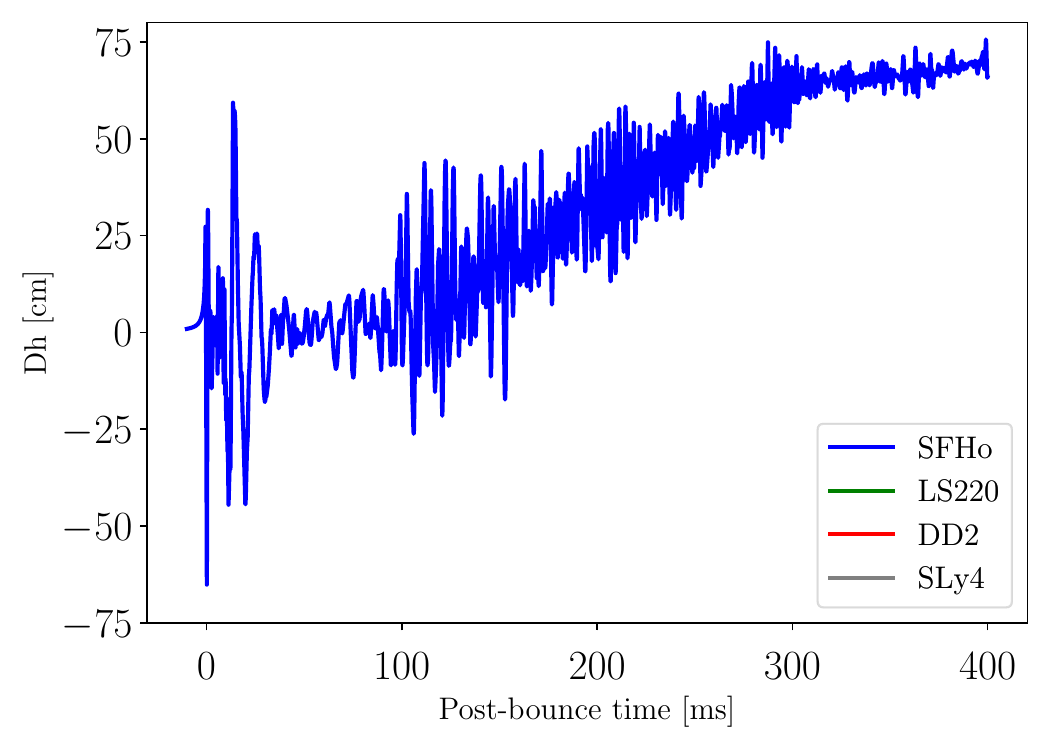}
        \includegraphics[width=\linewidth,clip,trim={0 2.0cm 0 2.4cm}]{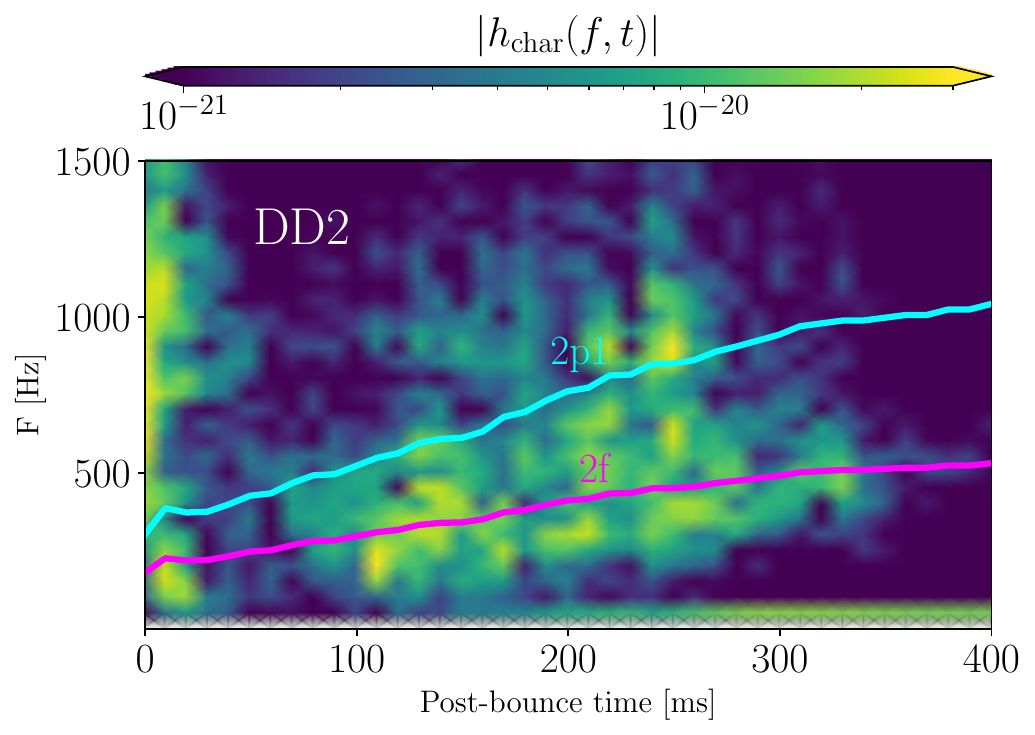}
        \includegraphics[width=0.97\linewidth,clip,trim={0 0 0 0cm}]{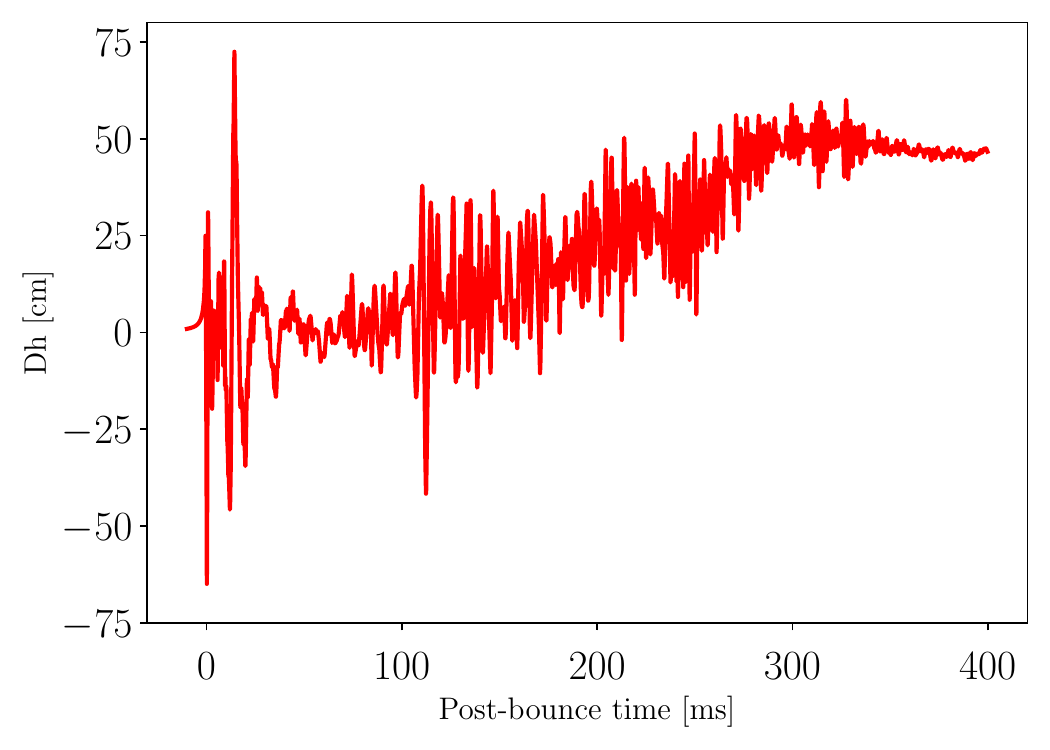}
    \end{minipage}%
    \begin{minipage}[t]{0.45\textwidth}
        \centering
        \includegraphics[width=\linewidth,clip,trim={0 2.0cm 0 0}]{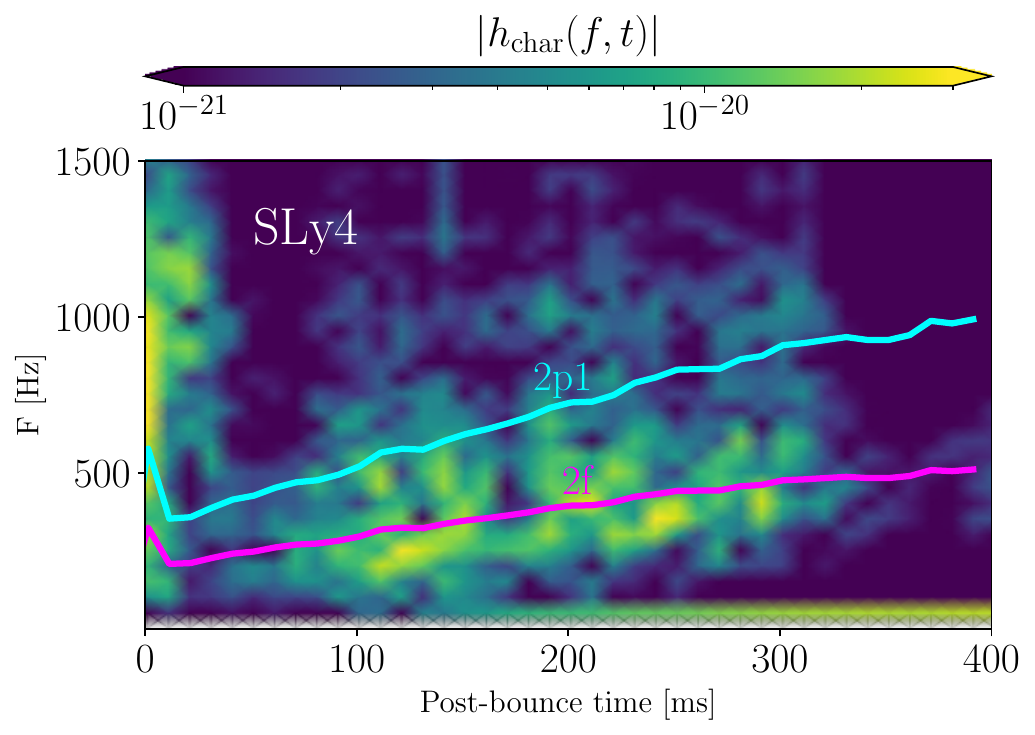}
        \includegraphics[width=0.97\linewidth,clip,trim={0 1.66cm 0 0cm}]{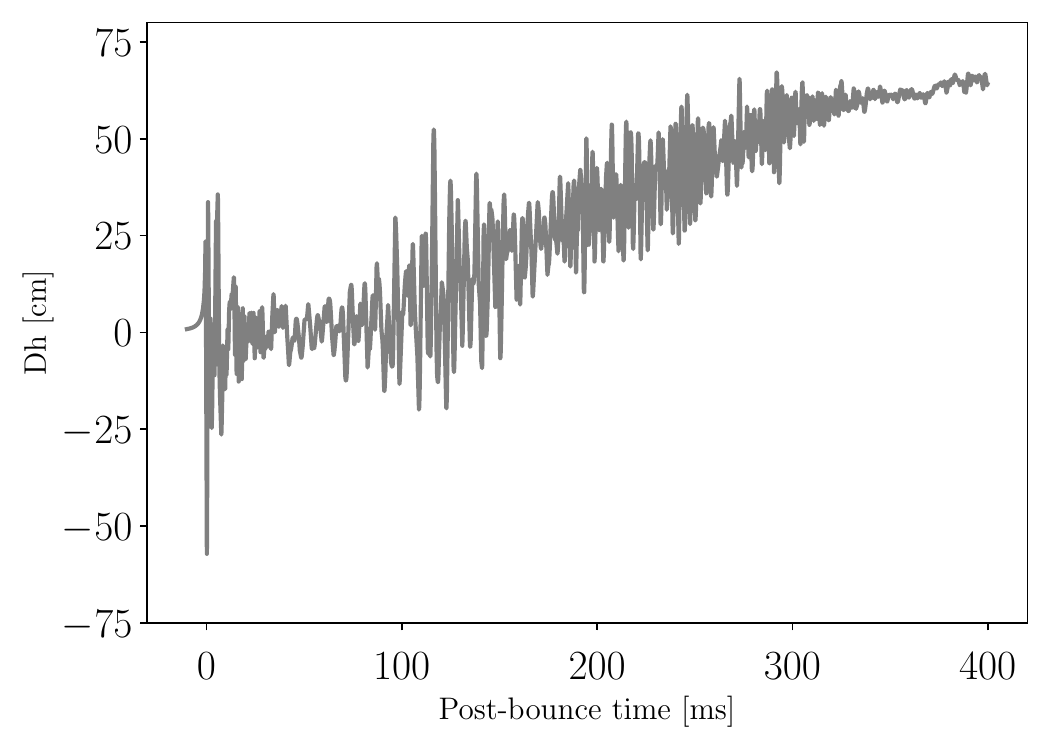}
        \includegraphics[width=\linewidth,clip,trim={0 2.0cm 0 2.4cm}]{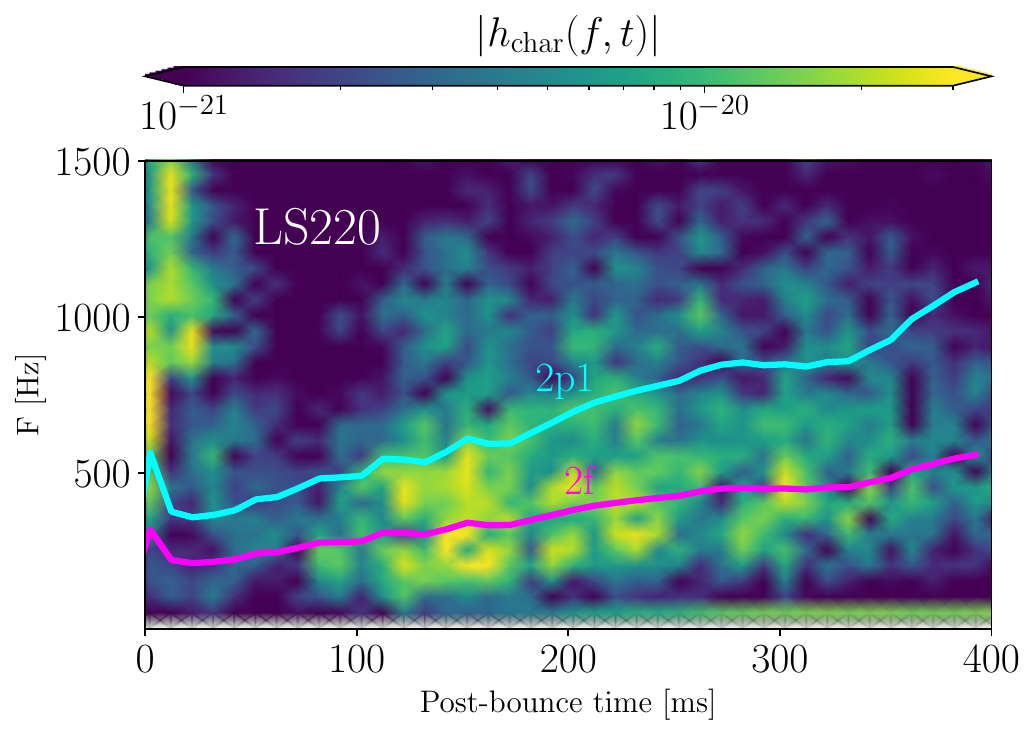}
        \includegraphics[width=0.97\linewidth,clip,trim={0 0 0 0cm}]{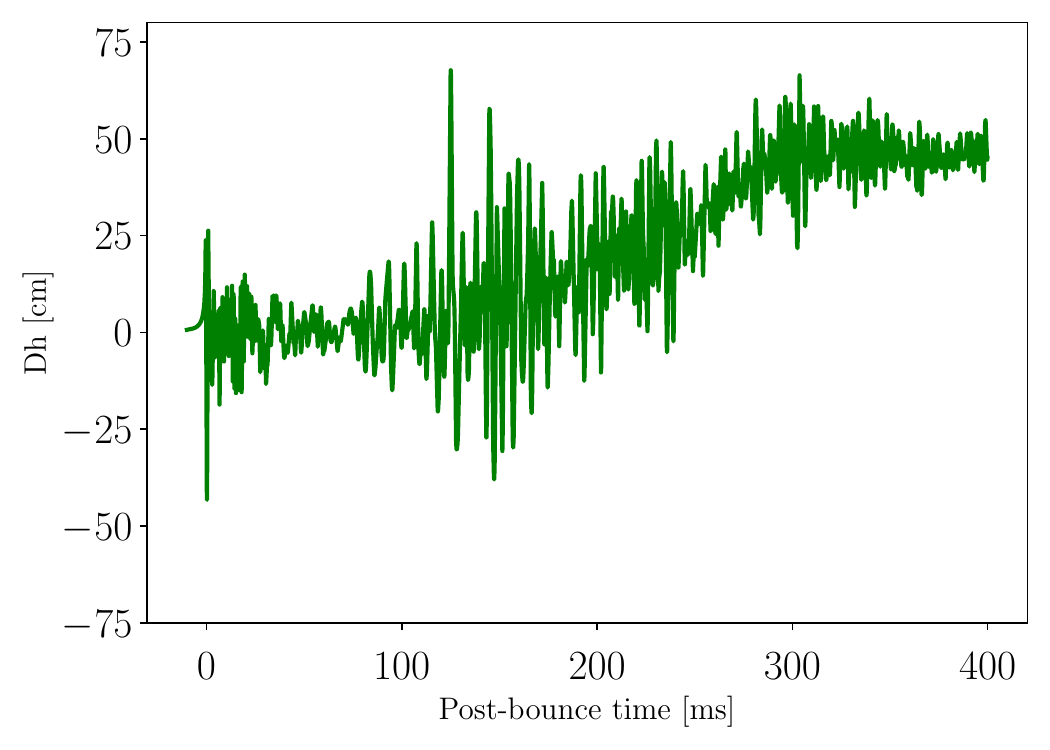}
    \end{minipage}
    \caption{GW signals for the four CCSN models. Each column corresponds to a different model: SFHo (top-left), DD2 (bottom-left) and SLy4(top-right), LS220 (bottom-right). In each column, the top panel shows the spectrogram of the gravitational-wave signal, highlighting the frequency evolution over time, while the bottom panel shows the corresponding time-series of the GW amplitude. This arrangement allows direct comparison of both the temporal and spectral features of the GW emission across different equations of state. 
    The pink line indicates the universal relation for the $2f$ mode, while the light-blue line shows the $2p_1$ mode.
}
    \label{fig:spect}
\end{figure*}

Figure~\ref{fig:spect} shows the full GW signal over the entire duration of the simulations. The deviation from zero encodes the degree of asymmetry in the explosion, a feature known as memory, which arises from the asymmetric distribution of the ejecta. Notably, SFHo stands out as the model producing the most asymmetric explosion, as indicated by its signal's long-term offset from zero.
Additional and particularly insightful information is provided by the spectrograms of the gravitational-wave characteristic strain $\mathrm{h}_\mathrm{char}$ computed at $10\,\mathrm{kpc}$ following \citet{kurodaGravitationalWaveSignatures2014, bugliThreedimensionalCorecollapseSupernovae2023} (Fig.~\ref{fig:spect}), which reveal that the post-bounce behavior
differs significantly across the four EoS models.
The spectrogram of the gravitational wave strain is computed via the Short-Time Fourier Transform (STFT), with a sampling frequency of the input signal \( f_s=10^4 \) and a time window of $20\,\mathrm{ms}$.
In the figure, the pink and light-blue curves indicate the universal relations from \cite{torres-forneUniversalRelationsGravitationalwave2019} for the $^2f$ and $^2p_1$ modes, respectively.

The spectrograms in Fig.~\ref{fig:spect} reveal a complex time-frequency structure that can be more clearly interpreted by following the post-bounce evolution in time.
At core bounce, all models exhibit a high-frequency component around $\sim 1\text{--}1.2\,\mathrm{kHz}$, which is associated with the bounce itself and the rapid excitation of PNS oscillation modes. This feature is common to all EoSs and reflects the global dynamical response of the inner core to the bounce. Small differences in the peak frequency are observed among the models, with LS220 and SLy4 showing slightly lower values compared to SFHo and DD2. As discussed in \citet{Richers2017}, such variations are largely consistent with differences in the central density reached at bounce, since the characteristic frequency scales approximately with the dynamical frequency $\sqrt{G\rho_c}$, and therefore only indirectly reflects differences in the underlying EoS.

In the early post-bounce phase ($\sim 10$--$50\,\mathrm{ms}$), differences between models become more evident. SFHo, DD2, and SLy4 all show a low-frequency component associated with prompt convection behind the shock. In LS220 this contribution appears significantly weaker in amplitude, by roughly an order of magnitude compared to the other models, although still present in the spectrogram. This indicates a reduced efficiency or a shorter-lived convective phase in this case.

A particularly interesting result is the strong correlation between the prompt-convection gravitational-wave emission and the incompressibility parameter $K$ of the EoS (mid-right panel of Fig.~\ref{fig:correlations}). 
To quantify this feature, we compute the RMS strain amplitude in the time interval between $5$ and $50\,\mathrm{ms}$ post-bounce and in the frequency range between $100$ and $300\,\mathrm{Hz}$, corresponding to the prompt-convection component identified in the spectrogram. The RMS measure provides a proxy for the typical GW amplitude in this band, and is not affected by the sign oscillations of the strain.
We find a remarkably strong positive correlation, with a Pearson coefficient $R = 0.994$, indicating that models characterized by a larger incompressibility systematically produce a stronger prompt-convection GW signal.
This trend is consistent with the dependence of the early post-shock dynamics on the stiffness of the EoS, which affects the strength and development of prompt convective motions immediately after bounce.

At later times ($\gtrsim 100\,\mathrm{ms}$ post-bounce), the GW signal is dominated by PNS oscillation modes. All models exhibit a clear ``ramping-up'' feature corresponding to a gradual increase of the dominant emission frequency from $\sim 100$ to $\sim 500\,\mathrm{Hz}$, which can be associated with the evolution of the PNS $f$-mode as the star contracts. 
Quantitatively, the rate of increase and the absolute frequency values show some dependence on the specific dynamical conditions. In particular, studies of non-rotating core-collapse models report a more rapid increase of the characteristic frequency, reaching values of order $\sim 1\,\mathrm{kHz}$ on similar timescales \citep[e.g.,][]{Jardine2022,torres-forneUniversalRelationsGravitationalwave2019}. In comparison, in our rapidly rotating models the frequency growth appears more moderate and remains typically below this range during the same post-bounce phase, as in the magnetorotational case shown in \citet[cf. Fig.~8, bottom panel]{Jardine2022}. This suggests that rotation plays a key role in modulating the evolution of the dominant PNS oscillation frequency by affecting the effective compactness evolution and mode structure of the remnant, rather than altering the qualitative behavior of the signal.
While this trend is present in all cases, its relative strength and persistence differ across models. In LS220, the emission associated with this ramping-up feature spans a broader frequency range at a given time, resulting in a wider low-frequency band. This broader distribution makes the frequency drift less sharply identifiable in LS220 compared to the other models.
At higher frequencies ($\gtrsim 1\,\mathrm{kHz}$), more pronounced differences emerge. DD2 exhibits the strongest late-time high-frequency emission, particularly around $\sim 250\,\mathrm{ms}$ post-bounce. The signal in this regime shows a broad-band structure rather than a set of clearly separated narrow modes. These components are consistent with pressure ($p$-) mode activity in the PNS, although this identification should be interpreted with caution, as the evolution appears to deviate from the behavior predicted by universal relations calibrated for neutrino-driven explosions \cite{torres-forneUniversalRelationsGravitationalwave2019} (here evaluated using the PNS mass and radius rather than the shock properties, and expressed as a function of $x=\sqrt{M_{\mathrm{PNS}}/R_{\mathrm{PNS}}^{3}}$. The dependence is fitted with a quadratic form $f = b x + c x^{2}$, with coefficients $b/10^5 = 1.410 \pm 0.004$, $c/10^6 = -4.23 \pm 0.06$ for $2f$, and $b/10^5 = 2.205 \pm 0.007$, $c/10^6 = 4.63 \pm 0.09$ for $2p_1$.). On the opposite end, SLy4 shows the weakest high-frequency emission among all models.
Another notable difference concerns the late-time behavior beyond $\sim 350\,\mathrm{ms}$ post-bounce. LS220 maintains a sustained GW emission across a broad frequency range, whereas in the other three models the signal is significantly suppressed at late times, except in a narrow region around the $f$-mode track.
These differences in the amplitude and persistence of the GW emission across the models likely reflect variations in the efficiency of mode excitation, which are in turn influenced by the PNS structure, the explosion dynamics, and the accretion dynamics.

\begin{figure}
    \centering
    \includegraphics[width=\linewidth]{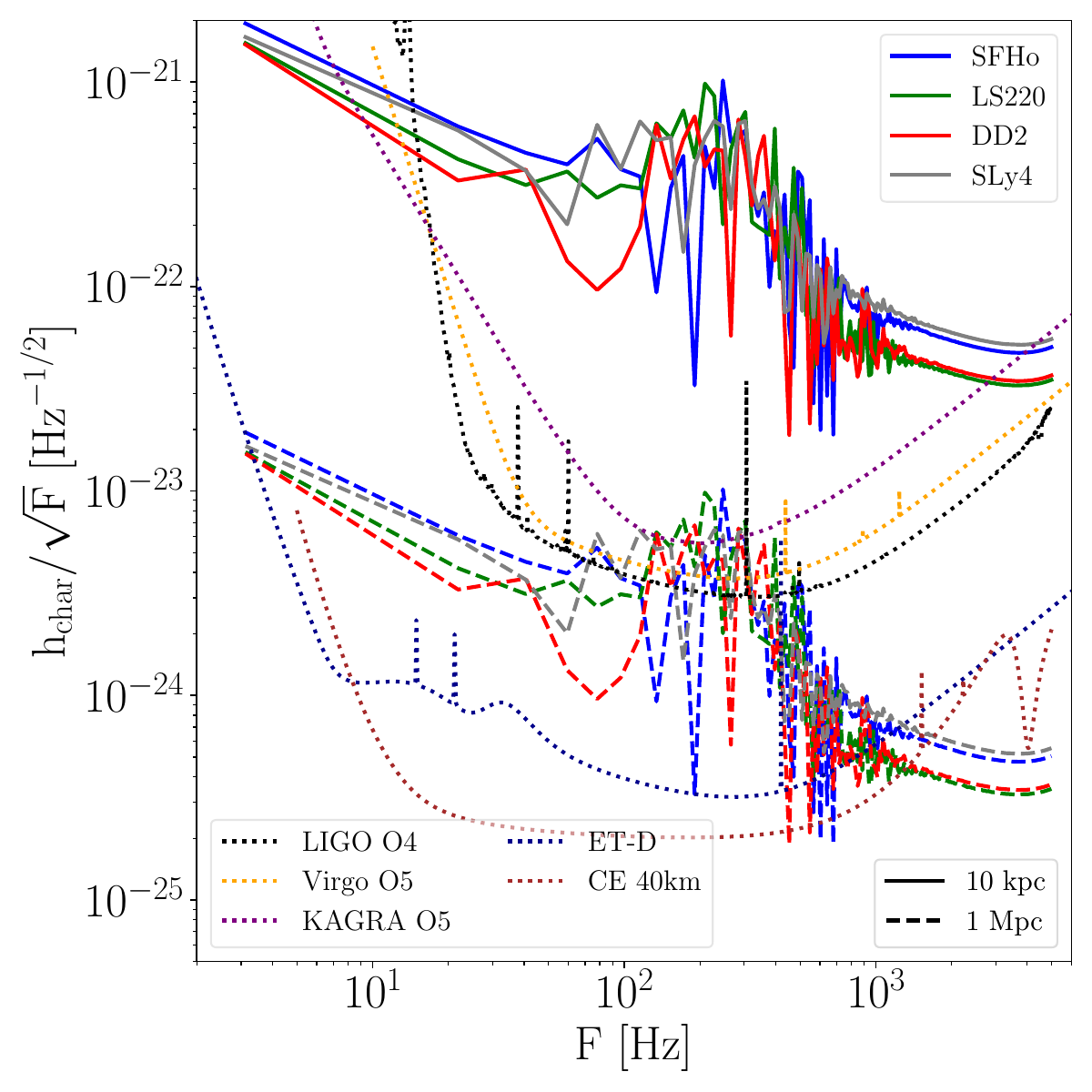}
    \caption{\small Time-integrated characteristic strain divided by the square root of frequency for all models, compared with the sensitivity curves of current (LIGO, Virgo, KAGRA; black, pink, dark blue) and future (Einstein Telescope and Cosmic Explorer with $40\,\mathrm{km}$ arms; yellow, brown, purple) gravitational-wave detectors. The strain is computed assuming source distances of $10\,\mathrm{kpc}$ (solid line) and $1\,\mathrm{Mpc}$ (dash-dotted line). The signals are smoothed using a first-order Butterworth filter with a cutoff frequency of 300 Hz.
    }

    \label{fig:sensitivity}
\end{figure}

\begin{table*}[t!]
    \centering
    \caption{Optimal SNR and maximum detection distance $D_\mathrm{max}$ [kpc] for different GW detectors and CCSN models ($D_\mathrm{ref} = 10$ kpc).}
    \begin{tabular}{lcccccccccccc}
        \hline
        & \multicolumn{2}{c}{\textbf{LIGO O4}} 
        & \multicolumn{2}{c}{\textbf{Virgo O5}} 
        & \multicolumn{2}{c}{\textbf{KAGRA O5}} 
        & \multicolumn{2}{c}{\textbf{ET-D}} 
        & \multicolumn{2}{c}{\textbf{CE 40 km}} \\
        \hline
        Model & SNR & $D_\mathrm{max}$  & SNR & $D_\mathrm{max}$ & SNR & $D_\mathrm{max}$ & SNR & $D_\mathrm{max}$ & SNR & $D_\mathrm{max}$ \\
        \hline
        SFHo  & 56.50 & 70.62 & 45.67 & 57.09 & 28.18 & 35.22 & 532.36 & 665.45 & 858.98 & 1073.73 \\
        LS220 & 75.95 & 94.94 & 62.26 & 77.82 & 39.89 & 49.87 & 724.81 & 906.01 & 1179.62 & 1474.53 \\
        DD2   & 54.44 & 68.05 & 44.16 & 55.20 & 27.54 & 34.42 & 514.86 & 643.57 & 843.08 & 1053.86 \\
        SLy4  & 52.24 & 65.31 & 42.58 & 53.23 & 26.69 & 33.36 & 495.96 & 619.95 & 807.09 & 1008.86 \\
        \hline
    \end{tabular}
    \label{tab:snr_dmax_all}
\end{table*}

Fig.~\ref{fig:sensitivity} shows the comparison between the time-integrated characteristic strain, obtained from the Fourier transform of the GW strain and expressed as a frequency-dependent amplitude weighted by the signal power, divided by the square root of the frequency, and the sensitivity curves of current gravitational-wave observatories (LIGO, Virgo, and KAGRA) as well as future detectors (the Einstein Telescope and the Cosmic Explorer, with $40\,\mathrm{km}$ arm lengths, respectively).
 It is particularly interesting to note that, at a distance of $10\,\mathrm{kpc}$, current observatories are already capable of detecting the signal in the $\sim 20\,\mathrm{Hz}$ to $1\,\mathrm{kHz}$ frequency range at least, providing a potentially observable signal even with current detectors. This band encompasses the most dynamical phases of magnetorotational core collapse, including bounce and early post-bounce activity, where strong GW emission is expected. Detectability in this range therefore provides direct access to the underlying microphysics and to the magnetic and rotational structure of the collapsing core.

This comparison also highlights the crucial role of next-generation detectors, which will enable access to the full signal at $10\,\mathrm{kpc}$ and substantially extend the distance reach of such observations, as shown by the dashed line that is the signal at $1\,\mathrm{Mpc}$. Their improved low- and high-frequency performance will make it possible to probe the detailed time–frequency evolution of the explosion mechanism and to perform precise model discrimination. The broader frequency coverage of future facilities is particularly relevant in view of the significant model-to-model variability introduced by uncertainties in the nuclear EoS, rotation rate, and magnetic-field topology. Enhanced sensitivity will help disentangle these effects by enabling a more complete reconstruction of the GW spectrum.
This will also strengthen the potential of data-driven inference techniques, including recent machine-learning approaches aimed at constraining the nuclear EoS directly from CCSN GW observations \citep{MoreRealisticMachinelearning2026,mitraGeneralizationGapMachine2026}.

In Table~\ref{tab:snr_dmax_all} we report the optimal signal-to-noise ratio (SNR) and the corresponding maximum detection distance $D_\mathrm{max}$ for the CCSN models considered. 
The SNR is computed at a reference distance of $D_\mathrm{ref} = 10$ kpc using the frequency-domain GW signal $h(f)$ as
\begin{equation}
    \mathrm{SNR} = \left[ 4 \int_0^\infty \frac{| \tilde{h}(f) |^2}{S_n(f)} \, \mathrm{d}f \right]^{1/2},
\end{equation}
where $S_n(f)$ is the one-sided noise power spectral density of the detector. 
To reduce spectral leakage associated with the finite duration of the numerical signal, we apply a Hann window to the time-domain waveform before performing the Fourier transform. This choice slightly reduces the total signal power but provides a more stable estimate of the frequency-domain signal entering the SNR computation. 
Here, “optimal SNR” refers to the matched-filter SNR obtained under the assumption of perfect template matching, and should be interpreted as an upper limit to the detectability of the signal.
 
Assuming that the SNR scales inversely with distance, the maximum detection distance is defined as the distance at which the signal reaches a threshold $\mathrm{SNR}=8$, i.e.
\begin{equation}
    D_\mathrm{max} = D_\mathrm{ref} \, \frac{\mathrm{SNR}(D_\mathrm{ref})}{8}.
\end{equation}

Each row of Table~\ref{tab:snr_dmax_all} corresponds to a different EoS model, while for each detector we report the SNR at $10$ kpc and the corresponding $D_\mathrm{max}$. 
This presentation highlights the improved sensitivity of next-generation detectors, which are capable of detecting signals from significantly larger distances compared to the current generation.

\subsection{$r$-process insight}
\begin{figure}
    \centering
    \includegraphics[width=\linewidth]{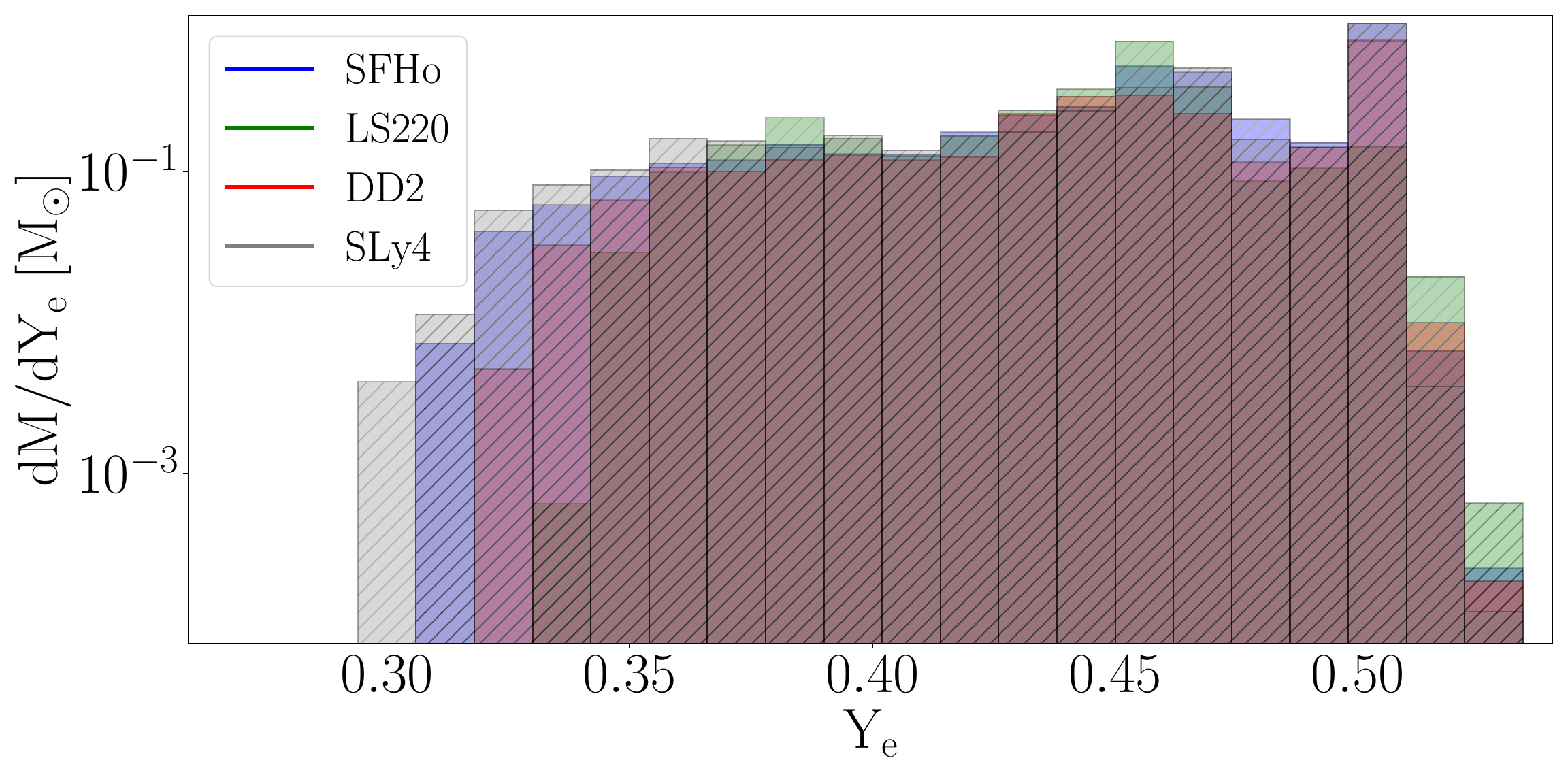}
    \caption{Mass fraction of neutron-rich matter ($\mathrm{Y}_{\mathrm{e}} \leq 0.5$) for the ejecta as a function of the electron fraction.}
    \label{fig:dMdY}
\end{figure}

In this subsection, we aim to provide some insight into the EoS on $r$-process nucleosynthesis in extreme CCSNe. 
The \texttt{Aenus-Alcar} code does not include a full nuclear reaction network, so we cannot compute the abundances of $r$-process elements directly. However, we can study the conditions in the ejecta, where $r$-process nucleosynthesis would occur. In particular, the shock temperature and the time evolution of the electron fraction provide useful diagnostics. Another important indicator is the mass fraction of the ejecta as a function of the electron fraction, which gives information about the neutron richness of the ejected material. The neutron richness directly affects the nucleosynthesis path, determining which heavy elements can be produced and their final abundances.

A faster shock revival tends to expel neutron-rich matter more rapidly, reducing the neutrino irradiation near the PNS and leading to lower electron fractions in the ejecta. Consistently, Fig.~\ref{fig:dMdY} shows that this trend results in more neutron-rich ejecta. Indeed, the SLy4 EoS is the only one that produces a significant portion of ejecta with $\mathrm{Y}_{\mathrm{e}} \sim 0.3$, while also yielding the smallest fraction of proton-rich ejecta ($\mathrm{Y}_{\mathrm{e}} \geq 0.5$). In contrast, the LS220 EoS produces a larger fraction of proton-rich ejecta and only a very small amount of material with $\mathrm{Y}_{\mathrm{e}} \sim 0.35$. These variations among different EoSs arise from their distinct nuclear compressibilities, symmetry energies, thermodynamic conditions throughout the core-collapse evolution, which influence the shock dynamics and the subsequent ejection of matter. Although a full nucleosynthesis calculation is beyond the scope of the present study, these results indicate which conditions are most promising for $r$-process production and can guide future simulations including detailed reaction networks.

\section{Conclusions}

\label{sect:conclusion}
We have investigated the impact of different nuclear EoSs on the explosion dynamics and multimessenger signals of magnetorotational core-collapse supernovae. Using a rotating $20\,\mathrm{M}_\odot$ progenitor, we performed axisymmetric simulations with spectral neutrino transport (12 energy bins), varying the EoS among SFHo, LS220, DD2, and SLy4.  

The EoSs differ not only in their parameterization (e.g., the values of incompressibility \( K \), symmetry energy \( S \), and its slope \( L \)) but also in the underlying physical models used to describe dense matter. While SFHo and DD2 are based on a RMF approach with meson exchange, they include a statistical treatment of non-homogeneous nuclear species matched to homogeneous nuclear matter, SLy4 and LS220 use a Hamiltonian framework based on Skyrme interactions, where nuclei are modeled as liquid drops with additional surface and Coulomb corrections.

Our results highlight that even EoSs with nominally similar nuclear parameters can produce qualitatively and quantitatively different outcomes. Key differences are observed in shock evolution, PNS radii, and moments of inertia. Softer EoS such as SLy4 are generally associated with faster shock expansion and more energetic ejecta, while LS220 is characterized by larger PNS radii, a broader neutrinosphere, and more sustained high-frequency gravitational-wave emission. SFHo and DD2 display intermediate behaviors and share similar low-frequency GW modes associated with prompt convection.  

An important aspect of the present study is that all models exhibit broadly similar explosion dynamics, characterized by magnetorotational shock revival on comparable timescales and a predominantly bipolar outflow geometry. As a consequence, variations in the PNS properties are less strongly influenced by differences in the global explosion dynamics than is typically the case in core-collapse simulations. This reduces the impact of model-to-model variations in explosion morphology and onset time, thereby allowing a clearer assessment of the role of EoS-dependent microphysical properties in shaping the late-time accretion history and PNS evolution.  

The gravitational-wave and neutrino signals are strongly EoS-dependent. LS220 produces strong, high-frequency, time-modulated GW emission from bounce-induced oscillations, while SLy4 shows a brief low-frequency signal associated with prompt convection. SFHo and DD2 exhibit more sustained low-frequency modes. Neutrino emission follows similar trends: SLy4 produces a sharp breakout burst due to its extended neutrinosphere, SFHo yields a broader, less intense emission, and LS220 shows the strongest late-time polar luminosity. These variations reflect the interplay between EoS microphysics, PNS structure, and explosion asymmetry.
Importantly, our results indicate that the nominal stiffness of an EoS alone does not fully explain these trends; a consistent interpretation also requires accounting for differences in the nuclear interaction model, the treatment of nuclear species, and the thermodynamic evolution of the system.

Regarding detectability, all models produce Galactic signals above the sensitivity of current GW detectors (LIGO, Virgo, KAGRA), while next-generation observatories (Einstein Telescope, Cosmic Explorer) would enable observations at larger distances. The distinctive features of each EoS in both GW frequency content and neutrino emission suggest that multimessenger observations could, in principle, help distinguish between different EoSs, although detailed 3D simulations will be required to fully quantify observational discriminants.

When compared to neutrino-driven explosions , we find that the matter responds differently in the two mechanisms. While all models in our magnetorotational simulations successfully explode, this is not necessarily the case in neutrino-driven explosions \citep{powell2025}. Notably, the evolution of PNS and neutrinosphere radii differs between the two mechanisms \citep{Steiner2013,suwaIMPORTANCEEQUATIONSTATE2013}, as do neutrino luminosities and mean energies \citep{Steiner2013}, and the morphology of the gravitational-wave signal \citep{Richers2017,murphyDependenceReconstructedCorecollapse2024}. These differences highlight that a comprehensive understanding of EoS effects requires extending studies to MHD explosions, as comparisons based solely on neutrino-driven models are insufficient.

Thermal effects in the EoS influence both the core-collapse dynamics and the structure of the PNS. 
A higher specific heat and thermal pressure can slow the contraction of the PNS, modify its oscillation frequencies, and affect the development of convection. 
Moreover, thermal differences during collapse and bounce affect the shock formation radius, which is systematically smaller in the LS220 case, and the subsequent shock evolution, which in turn influences the ejecta composition and the conditions for heavy-element formation via $r$-process nucleosynthesis.
These changes are reflected in the multimessenger signals: the amplitude and frequency content of gravitational waves, as well as the neutrino luminosities and spectra, are sensitive to the thermal properties of dense matter.

Previous studies provide useful complementary insight into the role of EoS microphysics in compact-object dynamics \citep[e.g.,][]{Yasin2018EoS,JankaBauswein2023}. 
In the context of neutrino-driven CCSNe, \citet{Fischer2014_EPJA} showed that $S$ significantly affects the collapse evolution, deleptonization, and the early post-bounce structure of the PNS. 
Although our simulations explore the magnetorotational regime, we find a consistent behavior in which $S$ correlates with several post-bounce observables, including the late-time PNS properties and the total neutrino emission.
This suggests that the influence of isospin-dependent microphysics on the collapse dynamics remains important independently of the explosion mechanism.
Similarly, in binary neutron-star merger simulations, \citet{Fields2023_ApJL} showed that thermal properties such as specific heat can significantly modify the dynamical response of dense matter and leave measurable imprints on the GW signal. 
Despite the different astrophysical context, our results point to an analogous role of thermal effects during core collapse. In particular, LS220 exhibits a longer collapse phase and shock formation at smaller radii, suggesting a different dynamical response during collapse. Conversely, the post-bounce evolution leads to the formation of the most extended PNS among the models considered.
This highlights that the compactness at shock formation and the late-time PNS structure is governed by different physical ingredients: the former is mainly set by collapse and bounce dynamics, while the latter emerges from the interplay of thermal support, rotation, accretion, neutrino cooling, and multidimensional effects.
Taken together, these comparisons support the broader picture that multimessenger observables in compact-object formation are sensitive not only to the cold stiffness of the EoS, but also to its thermal and compositional properties.

These conclusions are based on 2D axisymmetric simulations with approximate gravity, and on a single progenitor model. Future 3D studies with more accurate gravity treatments, a wider range of progenitors and rotation profiles, and additional microphysical ingredients (e.g., hybrid quark–hadron EoS) are required to robustly assess the impact of EoS on explosion dynamics, GW and neutrino emission, and nucleosynthesis.  

\begin{acknowledgements}
We thank T. Foglizzo, J. Guilet, and R. Raynaud for useful discussions.
We acknowledge CINECA for the availability of HPC resources through a CINECA–INFN agreement (allocation INF25\_teongrav).
MB acknowledges the support of the French Agence Nationale de la Recherche (ANR), under grant ANR-24-ERCS-0006 (project BlackJET), and the support of the European Union by the ERC grant BlackJET (n. 101164144). 
MC and MO acknowledge support from grant PID2021-127495NB-I00 funded by MCIN/AEI/10.13039/501100011033 and the European Union, as well as from the Astrophysics and High Energy Physics programme of the Generalitat Valenciana (ASFAE/2022/026), funded by MCIN and the European Union NextGenerationEU (PRTR-C17.I1), and from the Prometeo excellence programme grant CIPROM/2022/13 funded by the Generalitat Valenciana. 
MC acknowledges support by the Generalitat Valenciana via the grant CIDEGENT/2019/031.
Funded by the European Union. 
While partially funded by the European Union, views and opinions expressed are, however, those of the authors only and do not necessarily reflect those of the European Union or the European Research Council Executive Agency. 
Neither the European Union nor the granting authority can be held responsible for them.
\end{acknowledgements}

\nocite{*}
\bibliographystyle{aa}
\bibliography{references}

\begin{appendix}

\section{\texttt{Aenus-Alcar} equations}
\label{App:eqs}

\subsection{Fluid equations}

The code solves the set of hyperbolic RMHD equations for matter \citep{Obergaulinger2020} in conservative form:
\begin{align}
&\frac{\partial \rho_*}{\partial t} + \vec{\nabla} \cdot (\alpha \rho_* \vec{v}) = 0, \\
&\frac{\partial \vec{S}}{\partial t} + \vec{\nabla} \cdot 
\left[ 
    \alpha \left( 
        w_\mathrm{tot}\Gamma^2 \vec{v}\vec{v} 
        - \vec{b}\vec{b} 
        + p_\mathrm{tot}\,\mathbb{I} 
    \right)
\right] 
= \alpha \vec{Q}_M - \rho_* \vec{\nabla}\alpha, \\ 
&\frac{\partial U}{\partial t} + \vec{\nabla} \cdot (\alpha \vec{S}) 
= \alpha Q_E + \alpha\,\vec{v}\cdot\vec{Q}_M - \vec{S}\cdot\vec{\nabla}\alpha, \\
&\frac{\partial \vec{B}}{\partial t} + \vec{\nabla} \cdot 
\left[ 
    \alpha(\vec{B}\vec{v} - \vec{v}\vec{B})
\right] = 0, \\
&\frac{\partial (\rho_* Y_e)}{\partial t} + \vec{\nabla} \cdot (\alpha \rho_* Y_e \vec{v}) 
= \alpha Q_N.
\label{eq:RMHD}
\end{align}
Here, the operator $\nabla_i$ ($i = 1, 2, 3$) contains the determinant of the spatial metric, $\gamma$, which does not depend on time, $\Gamma$ is the Lorentz factor, $\rho_*$ is the mass density in the laboratory frame, $\vec{S}$ is the total momentum density, and $U$ is the total energy density.  Moreover, we define $w_\mathrm{tot}=\rho h + b^2$, $p_\mathrm{tot}=p+b^2/2$, with $\rho h = e+p$ being the specific enthalpy, where
\begin{equation}
 \vec{b} = \frac{\vec{B}}{\Gamma} + \Gamma(\vec{v}\cdot\vec{B})\vec{v}
    \label{eq:b four vector}
\end{equation}
is the comoving magnetic field. We also define
\begin{equation}
 b^0 = \Gamma(\vec{v}\cdot\vec{B}), \qquad b^2 = |\vec{b}|^2 - (b^0)^2 = B^2/\Gamma^2 + (\vec{v}\cdot\vec{B})^2.
    \label{eq:b four vector2}
\end{equation}
Finally, the conserved variables can be expressed in terms of the primitive variables as
\begin{align}
    \rho_* &= \rho \Gamma,\\
    \vec{S} &= (\rho h \Gamma^2 + B^2)\vec{v} - (\vec{v} \cdot \vec{B})\vec{B} 
    = w_{\mathrm{tot}} \Gamma^2 \vec{v} - b^0 \vec{b}, \\
    U &= \rho h \Gamma^2 - p + \frac{B^2}{2} + \frac{1}{2} \left( |\vec{v}|^2 |\vec{B}|^2 - (\vec{v} \cdot \vec{B})^2 \right) \nonumber \\
   & = w_{\mathrm{tot}} \Gamma^2 - (b^0)^2 - p_{\mathrm{tot}},
\end{align}
where the primitive variables are the rest mass density $\rho$, the fluid velocity $\vec{v}$, so that $\Gamma = (1-v^2)^{-1/2}$, the kinetic pressure $p$, and the magnetic field $\vec{B}$. For a more detailed discussion of these equations, see \cite{DelZanna2007, DelZanna2024, Mignone2024}. However, notice that, in addition to standard RMHD, here also the electron fraction $Y_e$ is evolved in time.

The other quantities that appear in the RMHD equations are the lapse function, $\alpha$, and the source terms that couple the fluid and neutrinos. These source terms describe the exchange of lepton number, momentum, and energy, denoted by $Q_N$, $Q_E$, and $\vec{Q}_M$, respectively, and are described in the next subsection.  The lapse function is computed from the gravitational potential $\Phi$, since the code does not implement a full $3+1$ treatment, so that \citep{Shapiro1983,OConnorOtt2010}
\begin{equation}
\alpha = \exp(\Phi), \qquad \vec{\nabla} \alpha = \alpha \vec{\nabla} \,\Phi,
\end{equation}
where the $\Phi$ is computed according to Case 'A' described in \cite{Marek2006EffectivePotential}.

\subsection{Neutrino equations}
\label{sub:neu_trans}
Neutrino transport is computed using the first two moments of the Boltzmann equation \citep{MunierWeaver1986,CernohorskyBludman_1994} with an maximum-entropy closure for the Eddington factor \citep{CernohorskyBludman_1994}, including the effects of gravity through the $\mathcal{O}(v)$-plus formulation described in \cite{endeveConservativeMomentEquations2012,Cardall2013}:
\begin{align}
    \begin{split}
         \partial_t E &+ \partial_t (v_i F^i) + \nabla_i [\alpha (F^i + v^i E) ]
    - (\nabla_i \alpha + \dot{v_i}) \left[ \partial_\epsilon (\epsilon F^i) - F^i \right] \\
  &  - \nabla_i (\alpha v_j) \left[ \partial_\epsilon (\epsilon P^{ij}) - P^{ij} \right] 
    = \alpha C^0,
    \label{eq:neutrino trasp mom 0}
    \end{split}\\
    \begin{split}
     \partial_t (F^i &+ v_j P^{ij}) + \nabla_{\!j} (\alpha P^{ij} + v^j F^i) + \dot{v^i} E 
    + \alpha F^j \nabla_{\!j} v^i  + (E + P) \nabla^i \alpha \\
    &  - \partial_\epsilon (\epsilon P^{i}_j) \dot{v^j}
    - \alpha \partial_\epsilon (\epsilon U_j^{ki}) \nabla_k v^j - \partial_\epsilon (\epsilon P^{ij}) \nabla_j \alpha = \alpha C^{(1),i},
    \label{eq:neutrino trasp mom 1}
    \end{split}
\end{align}
where $P^{ij}$ ($P=P^i_i$) and $U^{ki}_j$ are the second and third moments of the neutrino distribution function, respectively, and $\dot{\vec{v}}$ is the acceleration. 

The neutrino–matter interaction processes included in the simulation comprise nucleonic and nuclear scattering and absorption, inelastic electron scattering, electron–positron annihilation into neutrino–antineutrino pairs, and nucleon–nucleon bremsstrahlung. The source terms in Eq.~\ref{eq:RMHD} arising from neutrino–matter interactions can be defined from the collision integrals $C^{(0)}$ and $C^{(1),i}$ in Eqs.~\ref{eq:neutrino trasp mom 0} and \ref{eq:neutrino trasp mom 1}, respectively, as follows:
\begin{equation}
    Q_E = - \sum_{\text{species}} \bar{C}^{(0)},
    \label{eq:Q_E}
\end{equation}
\begin{equation}
    Q_M^i = - \frac{1}{c^2} \sum_{\text{species}} \bar{C}^{(1),i},
    \label{eq:Q_M}
\end{equation}
\begin{equation}
    Q_N = -m_B \int_0^{\infty} 
    \left[
        \left( \frac{C^{(0)}}{\epsilon} \right)_{\nu_e} 
        - 
        \left( \frac{C^{(0)}}{\epsilon} \right)_{\bar{\nu}_e}
    \right] 
    \mathrm{d}\epsilon.
    \label{eq:Q_N}
\end{equation}
For a more detailed description of these equations and their numerical implementation, see \cite{Just2015}.

\subsection{Gravitational-wave extraction}
\label{App:GW}

The GW signal is extracted using the standard axisymmetric quadrupole formalism.
Since our simulations do not evolve the spacetime metric, the GW emission is computed assuming a Newtonian matter source within the weak-field approximation, which is justified by the mildly relativistic conditions of the flow and the negligible contribution of higher-order relativistic terms.

Under axisymmetry, only the electric quadrupole mode $A^{E2}_{20}$ contributes to the signal. The observable strain is related to this quantity through
\begin{equation}
    h = \frac{1}{8}\sqrt{\frac{15}{\pi}} \frac{A^{E2}_{20}}{D},
\end{equation}
where $D$ is the source distance. Throughout this work, we therefore show the rescaled quantity $Dh$.

To optimize the numerical noise we compute the auxiliary quantity $N^{E2}_{20}$, obtained by replacing the first time derivative of the mass quadrupole moment through the continuity equation. The GW amplitude is then evaluated as
\begin{equation}
A^{E2}_{20} = \frac{d}{dt} N^{E2}_{20},
\end{equation}
where
\begin{equation}
\begin{split}
N^{E2}_{20}
=
\frac{G}{c^4}
\frac{32\pi^{3/2}}{\sqrt{15}}
\int_{-1}^{1}dz
\int_{0}^{\infty}dr \,
r^3 \rho
\bigg[
v_r \left(\frac{3}{2}z^2-\frac{1}{2}\right)
-
3 v_\theta z\sqrt{1-z^2}
\bigg].
\end{split}
\end{equation}
Here, $z=\cos\theta$, while $v_r$, and $v_\theta$ denote the velocity components in spherical coordinates. The remaining time derivative is evaluated explicitly from the simulation outputs, rather than using the momentum equation to eliminate the full second time derivative. Details on this entire procedure can be found in \citet{Moenchmeyer1991GWrot,Kotake2004, Obergaulinger2006}.

\section{Equation-of-state properties}
\label{app:EoS}

In this Appendix, we summarize several relevant properties of the considered equations of state that help interpret the differences observed in the collapse dynamics, PNS evolution, neutrino emission, and gravitational-wave signal. 
In particular, we discuss the density dependence of the effective adiabatic index and the mass-radius relations of cold, beta-equilibrated NS configurations.

\subsection{Effective adiabatic index profile}

\begin{figure}[b]
    \centering
    \includegraphics[width=\linewidth]{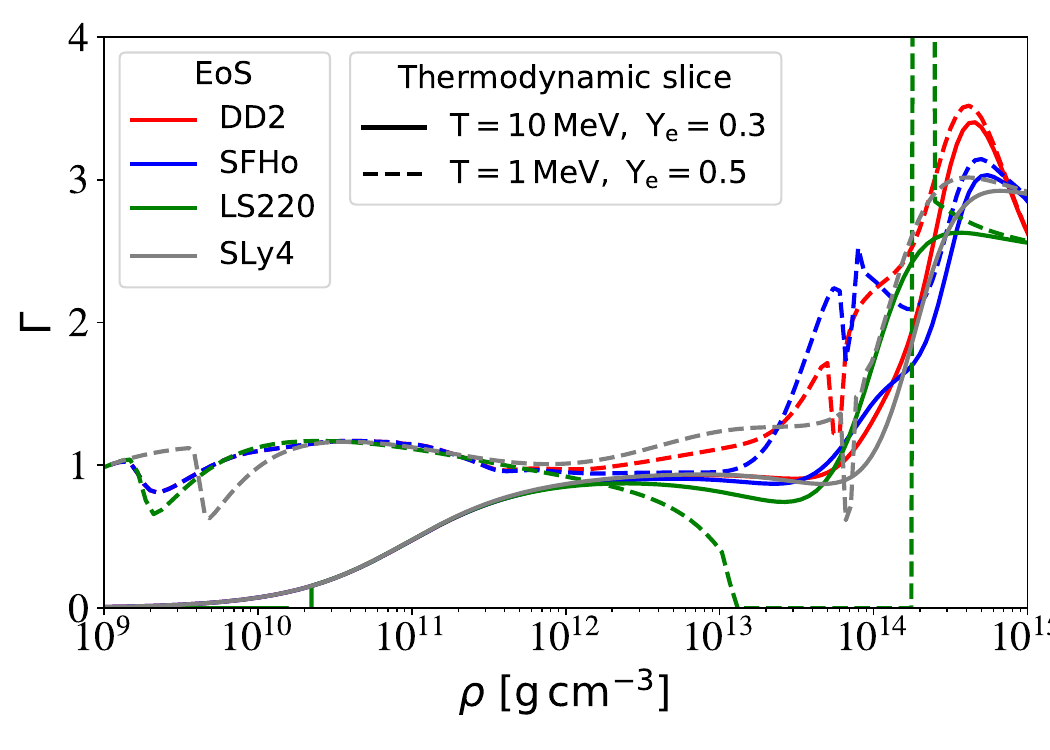}
    \caption{\small Effective adiabatic index as a function of density for the four considered EoSs, extracted from the \texttt{CompOSE} tables at two fixed thermodynamic conditions representative of core bounce ($T=10\,\mathrm{MeV}$ and $Y_e=0.3$) and of a less deleptonized, more symmetric composition and colder ($T=1\,\mathrm{MeV}$ and $Y_e=0.5$). Solid lines correspond to the former slice, while dashed lines indicate the latter. The profiles highlight the sensitivity of the effective thermodynamic response to both density and the underlying thermodynamic conditions.}
    \label{fig:gamma}
\end{figure}

To better characterize the thermodynamic response of the different EoSs during collapse, Fig.~\ref{fig:gamma} shows the effective adiabatic index as a function of density for two representative thermodynamic slices extracted from the tabulated EoS data. 
The first case corresponds to conditions representative of bounce in our simulations, namely $T=10\,\mathrm{MeV}$ and $Y_e=0.3$, while the second adopts $T=1\,\mathrm{MeV}$ and $Y_e=0.5$.

For the $T=10\,\mathrm{MeV}$, $Y_e=0.3$ slice, all EoSs display a qualitatively similar behavior. 
At low densities the effective adiabatic index is small and approaches values close to unity as the density increases, reflecting the relatively soft thermodynamic response of dilute matter dominated by nuclei, electrons, and thermal radiation. 
Around $\rho \sim 10^{12}\,\mathrm{g\,cm^{-3}}$ all models show a mild reduction, which is associated with the transition from non-uniform nuclear matter to increasingly homogeneous matter and with the progressive dissociation of heavy nuclei. 
This is followed by a gradual recovery and a second increase above $\sim 5\times10^{13}\,\mathrm{g\,cm^{-3}}$, as repulsive nuclear interactions become increasingly important approaching nuclear saturation density. 
Within this common trend, LS220 exhibits the most pronounced deviation, with a stronger dip and an anticipated recovery starting at lower densities ($\sim 3\times10^{13}\,\mathrm{g\,cm^{-3}}$), together with the lowest peak value at saturation, while DD2 shows the highest maximum. 
These differences reflect the different treatment of dense matter and thermal effects in the underlying EoS models, which modify the effective compressibility of the matter during the transition to nuclear densities.
A markedly different behavior emerges when switching to the $T=1\,\mathrm{MeV}$, $Y_e=0.5$ slice. 
In this case, at low densities all EoSs converge to $\Gamma \sim 1$, consistent with a colder and less pressure-supported regime, and remain more tightly clustered around $\rho \sim 10^{12}\,\mathrm{g\,cm^{-3}}$. 
However, the high-density behavior changes significantly, particularly for LS220. While DD2, SFHo, and SLy4 retain a qualitatively similar structure across the full density range—including a secondary reduction of $\Gamma$ around $\rho \sim 10^{14}\,\mathrm{g\,cm^{-3}}$—they systematically lie above the corresponding values obtained for the $T=10\,\mathrm{MeV}$, $Y_e=0.3$ slice and remain more closely grouped together.
In contrast, LS220 shows a much stronger deviation: the drop in $\Gamma$ above $\sim 10^{12}\,\mathrm{g\,cm^{-3}}$ is significantly more pronounced, and the evolution towards nuclear saturation is more abrupt, culminating in a sharp feature approaching a discontinuity at saturation density. 
This behavior suggests a stronger sensitivity of the effective compressibility to the thermodynamic state in LS220, likely connected to the specific treatment of the transition between non-uniform and homogeneous nuclear matter in this EoS.

\subsection{Mass-radius relation for NS equilibria}
\label{App:m-r}

\begin{figure}[t]
    \centering
\includegraphics[width=\linewidth]{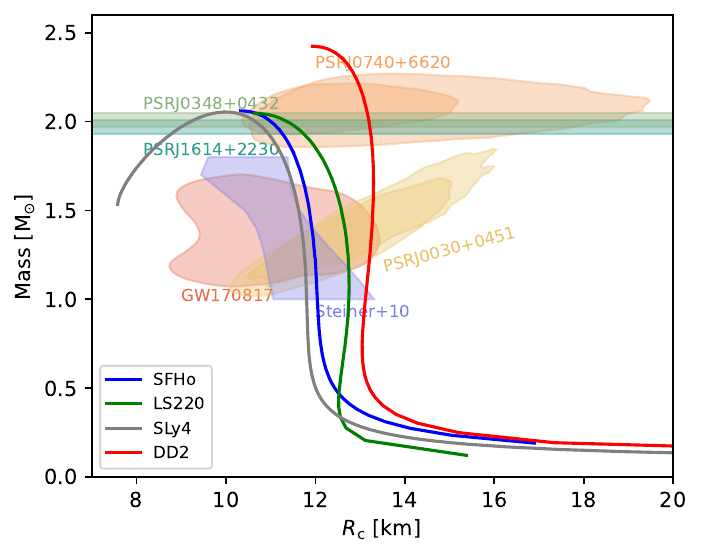}
\caption{Mass-radius relations for the EoS considered in this work, obtained from the \texttt{CompOSE} database. These curves represent cold NS hydrostatic equilibria (solving TOV equations) and provide a useful proxy for the stiffness of each EoS. Differences with the radii found in our simulations arise because PNS are hot, lepton-rich, and out of equilibrium. 
}

\label{fig:massvsradius}
\end{figure}

Here we present the mass–radius plots for cold, beta-equilibrated and isolated NS configurations obtained by using the four EoSs above described. The curves shown in Figure~\ref{fig:massvsradius} are obtained from the \texttt{CompOSE} database, based on solving the Tolman–Oppenheimer–Volkoff (TOV) equations, which describe hydrostatic equilibria in the framework of general relativity. Once an EoS is specified to close the TOV system, the stellar mass as a function of radius can be computed. The choice of EoS crucially affects the maximum mass that a PNS can support before collapsing into a black hole \citep{EspinoPaschalidis2019}. 

Comparing theoretical mass-radius relations with observational constraints provides a direct way to characterize the EoS in terms of macroscopic observables, rather than relying solely on microscopic nuclear-physics inputs. Therefore, our plot also contains data for the two most massive known pulsars, J0348$+$0432 \citep{Antoniadis2013} and J1614$-$2230 \citep{Demorest2010}, to prove that a chosen EoS can reach such limits. In addition to these precisely measured high-mass pulsars, we also include mass-radius constraints obtained from several observational channels. The colored regions correspond to the inferred mass and radius of a small NS population as derived by \citep{steinerEquationStateObserved2010}, shown at the $1\sigma$ (gray) and $2\sigma$ (green) confidence levels. To further constrain the mass-radius relation, we incorporate recent measurements from the NICER mission, which provides simultaneous mass and radius estimates through pulse-profile modeling of X-ray hotspots. In particular, we include the NICER analyses of PSR~J0030$+$0451 \citep{Miller19,Riley19} and PSR~J0740$+$6620 \citep{Miller21,Riley21}, which place stringent limits on the stellar compactness and thus on the underlying EoS. Additionally, we consider constraints from gravitational-wave observations of the binary NS merger GW170817 \citep{Abbott17}, where the tidal deformability measurements provide complementary limits on the pressure at supranuclear densities.

Overall, SFHo, DD2, and SLy4 are broadly consistent with current astrophysical and nuclear-matter constraints, while LS220, although still compatible with the astrophysical limits shown in Fig.~\ref{fig:massvsradius}, is generally disfavored by more recent nuclear-theory constraints and experimental determinations of dense-matter properties \citep[e.g.,][]{tewsSymmetryParameterConstraints2017}.

\section{Convective stability criterion}
\label{App:BV}

This appendix provides additional material on the convective stability analysis of the post-bounce configuration. In particular, we examine the thermodynamic gradients and the resulting stability properties in terms of the Brunt--V\"ais\"al\"a frequency, with the aim of identifying regions of convective instability within the PNS.

\subsection{Thermodynamic profiles and convective stability}
\begin{figure}
    \centering
\includegraphics[width=\linewidth]{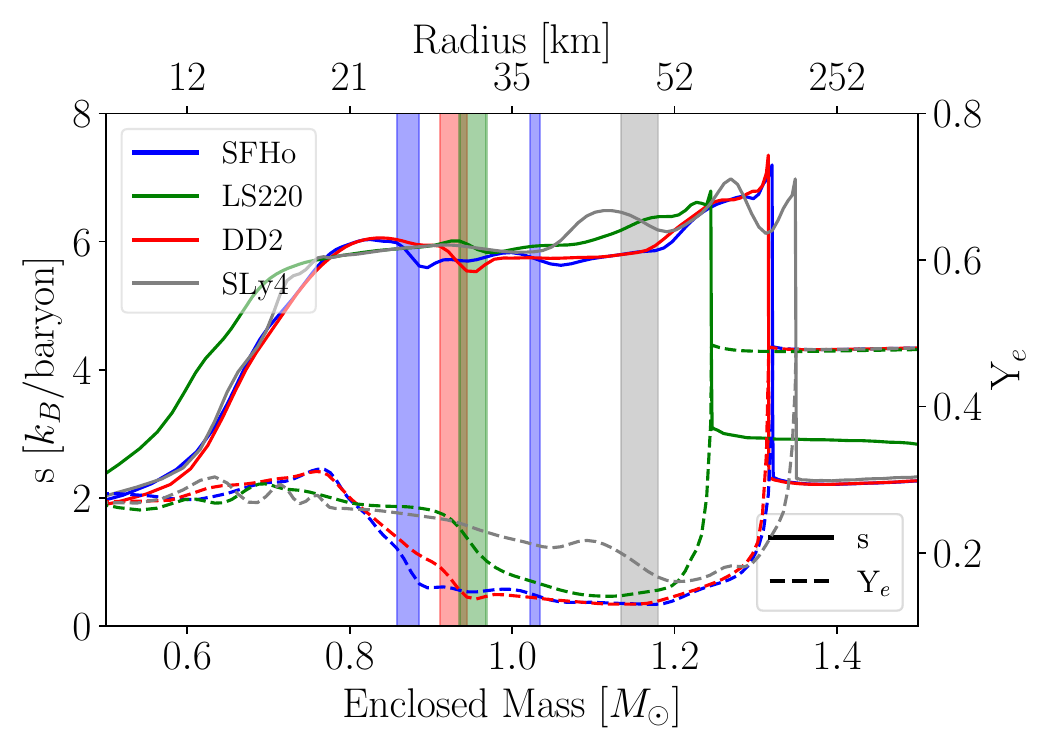}
\caption{Entropy (solid lines) and electron fraction (dashed lines) profiles as a function of enclosed mass at $t=10\,\mathrm{ms}$ post-bounce for the four EoSs. The upper axis shows the radial coordinate (in km) corresponding to the SFHo enclosed-mass profile and is provided as a representative reference for all models. The profiles are computed as an angular average over a narrow equatorial band ($\pm 5$ grid points around the equator, corresponding to an average in $\cos\theta$) to reduce numerical noise while preserving the intrinsically aspherical structure of the PNS. Shaded regions indicate convectively unstable zones, identified where both entropy and electron-fraction gradients are negative. These regions provide a proxy for the spatial extent and mass content of prompt convection.}

\label{fig:s_ye_mass}
\end{figure}

To identify the regions potentially contributing to prompt convection, Fig.~\ref{fig:s_ye_mass} shows the profiles of entropy and electron fraction as a function of enclosed mass at $t=10\,\mathrm{ms}$ post-bounce. The profiles are computed as an angular average over a narrow equatorial band ($\pm 5$ grid points around the equator, corresponding to an average in $\cos\theta$), in order to reduce numerical noise while preserving the intrinsically aspherical structure of the PNS. This choice is motivated by the strong departure from spherical symmetry, which makes a full spherical average less representative of the regions actually involved in convective activity.
Convectively unstable regions are identified by shaded bands where both entropy and electron-fraction gradients are negative. These regions provide a useful proxy for the spatial extent of prompt convection and the amount of mass effectively participating in it.

A clear qualitative difference emerges among the four EoSs. SFHo exhibits two distinct convectively unstable regions, one around $\sim0.9\,M_\odot$ and a second extending beyond $1\,M_\odot$, indicating multiple layers susceptible to prompt convection. This extended unstable structure is consistent with the relatively strong prompt convection signal observed in the corresponding gravitational-wave emission.
DD2 displays a single unstable region located at slightly higher enclosed mass compared to the first SFHo band, broadly similar to the unstable region found in LS220. However, in LS220 this region is slightly narrower in mass extent, which is consistent with a weaker convective driving and a correspondingly reduced GW amplitude.
SLy4 shows a qualitatively different behavior: the convectively unstable region is broader in mass but shifted toward larger enclosed masses, i.e. closer to the outer layers of the PNS. This more external location implies a weaker coupling to the inner quadrupolar dynamics of the core, which likely explains the reduced efficiency of GW emission despite the apparently extended unstable region. In this sense, the larger radial offset of the unstable layer can also contribute to a more gradual and slightly delayed development of the prompt convection signal compared to the other models.

\subsection{Brunt--V\"ais\"al\"a frequency}

\begin{figure}[]
    \centering
  \includegraphics[width=0.9\linewidth,clip,trim={0 1cm 0 0}]{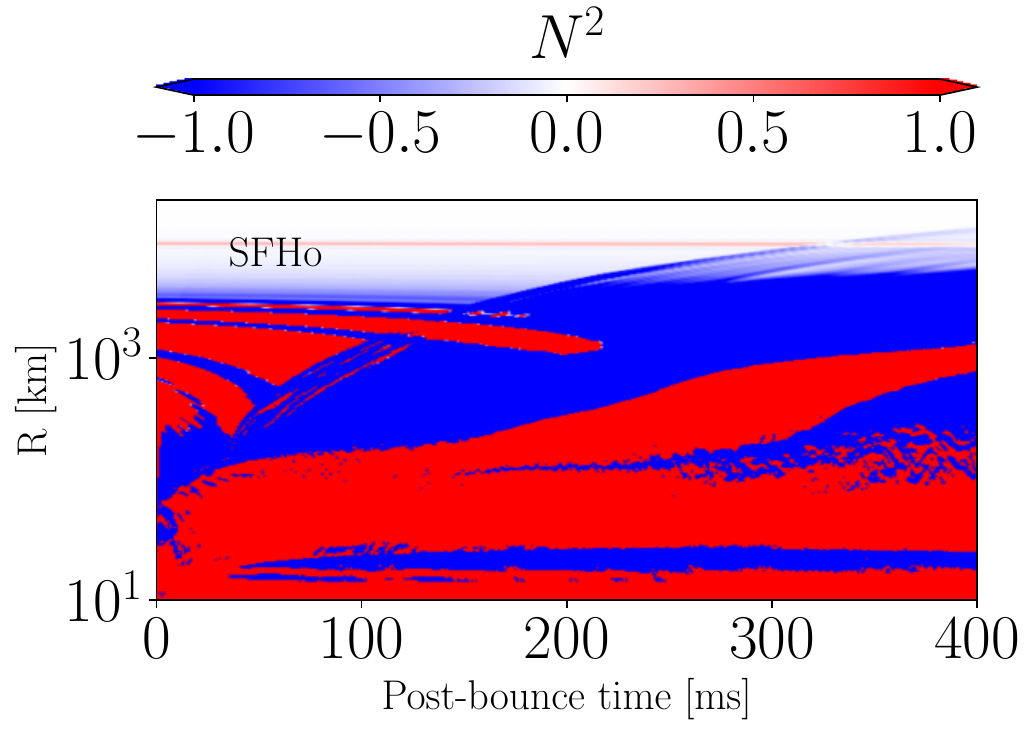}
    \includegraphics[width=0.9\linewidth,clip,trim={0 1cm 0 2.68cm}]{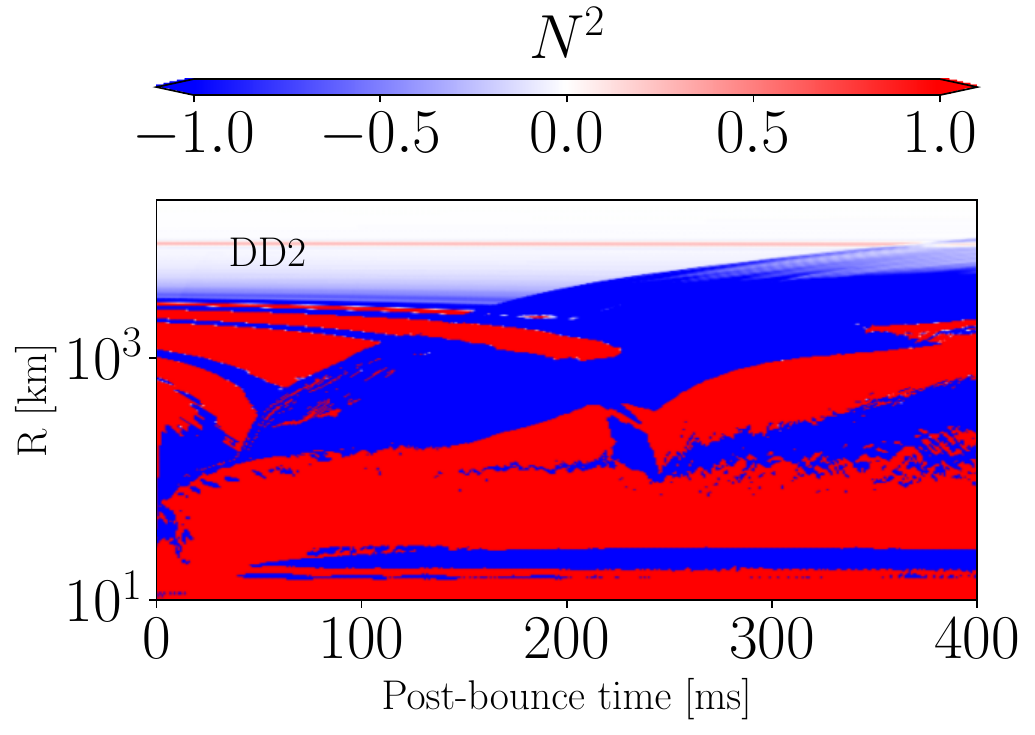}
    \includegraphics[width=0.9\linewidth,clip,trim={0 1cm 0 2.68cm}]{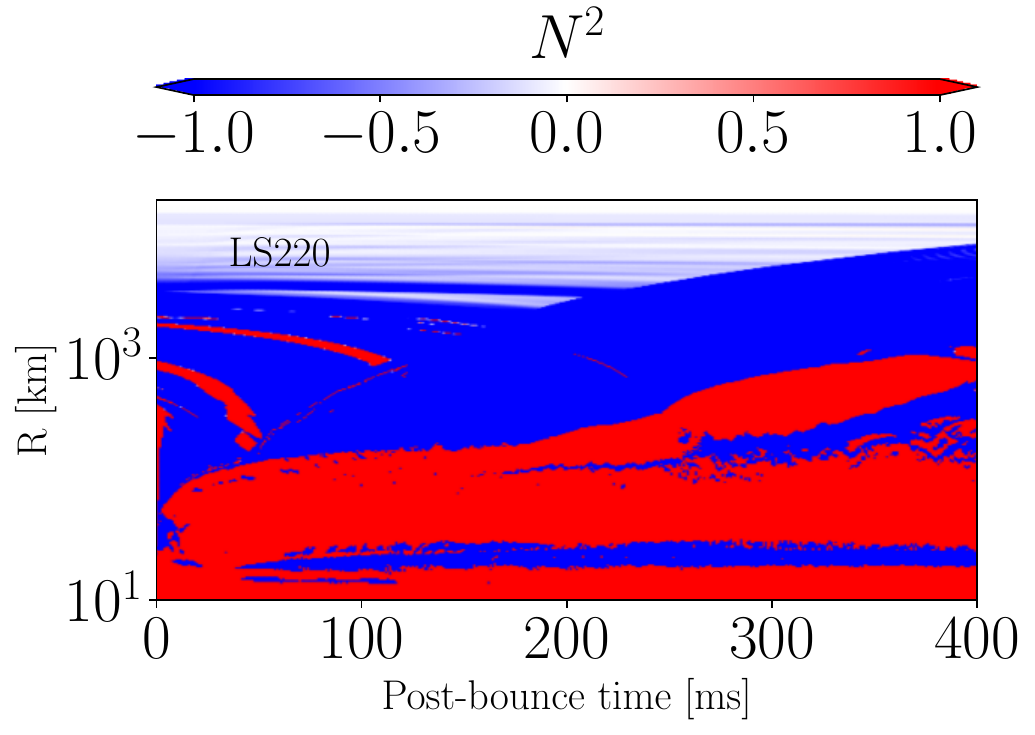}
    \includegraphics[width=0.9\linewidth,clip,trim={0 0 0 2.68cm}]{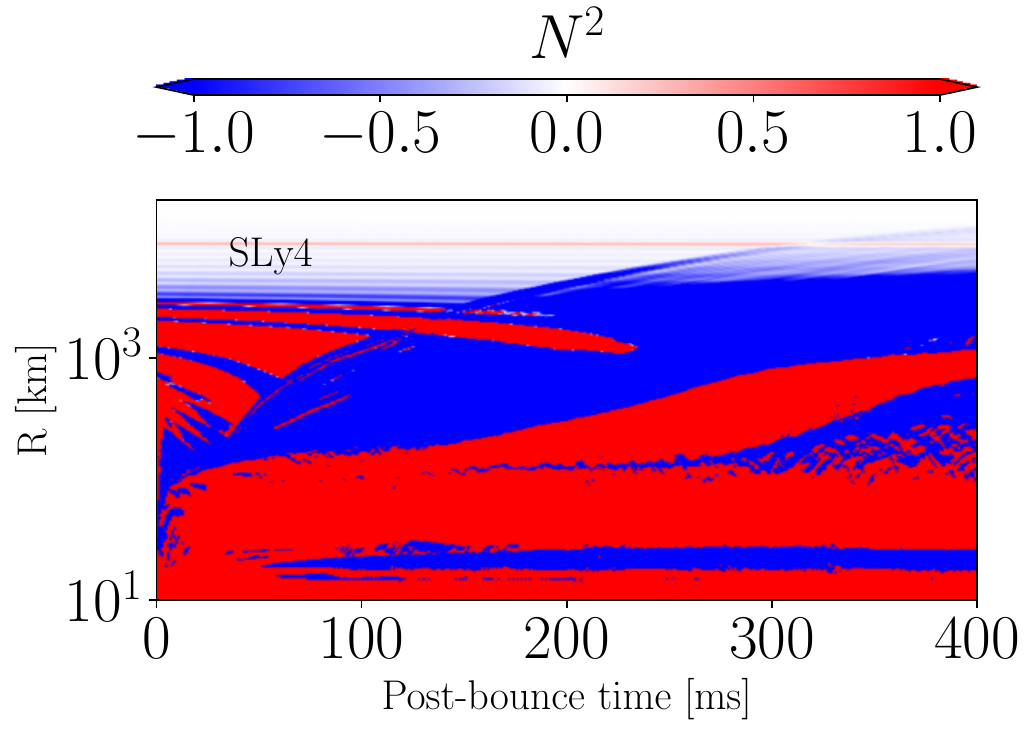}
\caption{\small Spacetime diagram of the Brunt-Väisälä frequency $N^2$ computed for our 4 models.}
\label{fig:BV}
\end{figure}

The sign of the Brunt-Väisälä frequency $N^2$ is known to trace the convective stability of the PNS. In Figure~\ref{fig:BV} we show, as a function of time and for all our 4 models, the sign of the Brunt–Väisälä frequency squared, $\mathrm{sign}(N^2)$, as a diagnostic of convective stability for all radii, averaged over the polar direction. The blue region indicates convectively unstable zones ($N^2<0$), while the red region corresponds to convectively stable ones ($N^2>0$) \citep{nagakuraSystematicStudyProtoneutron2020, SemiglobalSimulationsMagnetorotational}. The innermost part of the PNS is stable in all our models, whereas differences are observed in the prompt convection and in the Ledoux convection within the PNS.
For the LS220 model, prompt convection regions are visible immediately after bounce, while around $\sim 50\,\mathrm{ms}$ post-bounce Ledoux convection develops within the PNS. Among the other three EoSs, SLy4 and SFHo display similar levels of convective kinetic energy and corresponding $N^2$ patterns, whereas DD2 shows slightly lower convective activity, consistent with the weaker gradients in entropy and lepton number. 
In addition, all models show the development of Ledoux convection below the shock.

We note that the Brunt–Väisälä frequency $N^2$ provides a linear stability diagnostic based on the full thermodynamic response of the fluid, and is therefore a less restrictive criterion compared to our proxy based on simultaneous negative gradients of entropy and electron fraction. In particular, while $N^2<0$ captures general convective instability including composition- and entropy-driven contributions, the $(\partial s/\partial r < 0, \partial Y_e/\partial r < 0)$ condition isolates regions of strongly Ledoux-unstable stratification.
As a consequence, the two diagnostics should not be expected to provide identical spatial distributions of unstable regions, but rather to highlight different aspects of the convective stability of the post-bounce flow.

\end{appendix}

\end{document}